\documentclass[a4paper,amsmath,twocolumn,prd,nofootinbib,showpacs]{revtex4}
\usepackage{graphicx}
\usepackage{psfrag}
\usepackage[colorlinks=true,linkcolor=blue,citecolor=blue,urlcolor=blue]{hyperref}

\newcommand{\dd}{{d}}
\newcommand{\ee}{{e}}
\newcommand{\ii}{{i}}
\newcommand{\sign}{{\rm sign}}

\newcommand{\pdx}{\partial_x}
\newcommand{\pdt}{\partial_t}

\newcommand{\dom}{\dd\omega}

\newcommand{\php}[1]{\ee^{+\ii #1}}
\newcommand{\phm}[1]{\ee^{-\ii #1}}

\newcommand{\chor}{{c_0}}

\newcommand{\ka}{{\kappa_{\rm B}}}
\newcommand{\kk}{{\kappa_{\rm K}}}
\newcommand{\wB}{{w_{\rm B}}}
\newcommand{\kt}{{\kappa_{p}}}
\newcommand{\Dt}{{D_{p}}}
\newcommand{\wt}{{w_{p}}}
\newcommand{\nt}{{n_{p}}}

\newcommand{\wtl}{w_{{p}\delta}}

\newcommand{\dxl}{{d_\xi^{L}}}
\newcommand{\dxr}{{d_\xi^{R}}}

\newcommand{\Th}{{T_{\rm H}}}
\newcommand{\TB}{{T_{\rm B}}}

\newcommand{\om}{\omega}
\newcommand{\omok}{\omega/\ka}
\newcommand{\omm}{\omega_{\rm max}}
\newcommand{\ommax}{\omm}
\newcommand{\ommok}{\omega_{\rm max}/\ka}
\newcommand{\ommt}{\om_{{\rm Mp}}}

\newcommand{\run}{\Delta_{T}({\om})}

\newcommand{\runT}{\Delta_{T}({T_0})}

\newcommand{\pom}{\phi_\omega^\alpha(x)}

\newcommand{\vpom}{\varphi_\omega^\alpha(x)}
\newcommand{\vpoms}{(\varphi_\omega^{\alpha}(x))^*}

\newcommand{\aom}{\hat a_\omega^\alpha}
\newcommand{\aomd}{\hat a_\omega^{\alpha\dagger}}

\newcommand{\hP}{\hat\Psi}
\newcommand{\Pn}{\Psi_0}
\newcommand{\rn}{\rho_0}

\newcommand{\hp}{\hat\phi}

\newcommand{\hpd}{\hat\phi^\dagger}

\newcommand{\ha}{\hat a}

\newcommand{\dcrit}{D_{p}^{\rm crit}}

\begin{document}

\title{Spectral properties of acoustic black hole radiation:
Broadening the horizon}

\author{Stefano Finazzi}
\email{finazzi@sissa.it}
\affiliation{SISSA, via Bonomea 265, Trieste 34151, Italy and INFN sezione di Trieste, Via Valerio 2, Trieste 34127, Italy}
\author{Renaud Parentani}
\email{renaud.parentani@th.u-psud.fr}
\affiliation{Laboratoire de Physique Th\'eorique, CNRS UMR 8627, B{\^{a}}timent 210, Universit\'e Paris-Sud 11, 91405 Orsay Cedex, France}

\begin{abstract}
The sensitivity of the black hole spectrum when introducing short distance dispersion is studied in the context of atomic Bose condensates. By considering flows characterized by several length scales, we show that, while the spectrum remains remarkably Planckian, the temperature is no longer fixed by the surface gravity.
Rather it is determined by the average of the flow gradient across the horizon over an interval fixed by the healing length and the surface gravity, as if the horizon were broadened.
This remains valid as long as the flow does not induce nonadiabatic effects that produce oscillations or some parametric amplification of the flux.
\end{abstract}

\date{\today}
\maketitle

\section{Introduction}
\label{sec:intro}

In the hydrodynamic approximation, i.e. for long wavelengths, 
the propagation of sound waves in a moving fluid that crosses the speed of sound
is analogous to that of light in a black hole 
metric~\cite{Unruh81}. 
However, 
Hawking radiation relies on 
short wavelength modes~\cite{TJ91,93,Primer}. 
Therefore 
the 
dispersion of sound waves 
must be taken into account
when computing the spectrum 
emitted by an acoustic black hole. 
To this end, 
Unruh 
wrote 
a 
dispersive wave equation 
in a supersonic flow~\cite{Unruh95}. 
Through a numerical analysis, he then showed 
that the spectrum 
was robust, 
provided the dispersive 
scale 
$\xi$, the "healing length", %
is much smaller than the surface gravity scale $1/\kappa$ which fixes the 
Hawking temperature $\Th = \kappa/2\pi$
(in units where $\hbar = k_{\rm B} = c = 1$). 
This insensitivity 
was then confirmed by analytical~\cite{BMPS95,Corley97,Tanaka99,UnruhSchu05}
and numerical~\cite{CJ96,UnruhTrieste,MacherRP1,MacherBEC} methods. 
However it turned out that 
the {\it deviations} with respect to the standard 
flux are much harder to characterize, and
so far there is no consensus on what are the relevant parameters. 

To address this question, 
we consider 
velocity profiles 
characterized by {several} scales, and numerically 
compute the 
spectrum by varying them separately. 
Whenever $\omm/\TB \gtrsim 10$, where $\omm$ is a critical frequency~\cite{MacherRP1,MacherBEC} that 
scales with $1/\xi$ 
but also depends on the asymptotic flow velocity, 
the flux remains Planckian to a high accuracy, {\it even} 
when the temperature completely differs from $\Th$.
In fact, 
we show that 
the temperature is determined by the 
average of the velocity gradient 
across the 
horizon over a 
critical length 
 fixed by $\xi$ and $\kappa$.
In~\cite{ACRP2}, 
we analytically explain 
the origin of that length. 

These properties  hold 
for large classes of flows. 
They cease to be valid 
only if the flow induces 
nonadiabatic effects that interfere with the Hawking effect, thereby introducing 
oscillations in the spectrum.
This is illustrated by considering 
{\it undulations} similar to those appearing 
in {white} holes' flows~\cite{Carusottowhite,Silke}.
For definiteness we 
use the Bogoliubov--de~Gennes (BdG) equation 
in the context of atomic Bose condensates.
However our results apply to other analogue models, e.g. with subluminal dispersion relations,
and, more generally, to all dispersive theories.
 
We have organized the paper as follows. We first present the multiparameter 
flow profiles we shall use. We then 
analyze the phonon spectrum 
obtained by numerically solving the mode 
equation. 
In Appendixes we briefly present 
the concepts which are needed to obtain the phonon flux, starting
from the BdG 
equation.

\section{Choice of flow profiles}
\label{sec:bhlasers}

In this paper we consider 
elongated condensates that are stationary flowing along the 
longitudinal direction $x$. We also assume that the 
transverse dimensions 
are small enough that relevant 
phonon excitations are 
 longitudinal. In that case, 
one effectively deals with a 1+1 dimensional field.
Then, 
using the analogy~\cite{Unruh81,lr} between 
sound 
propagation 
in the hydrodynamic regime 
and 
light, 
the flow 
defines 
a metric
of the form 
\begin{equation}\label{eq:metric}
 \dd s^2=-c^2 \dd t^2+(\dd x-v \, \dd t)^2~,
\end{equation}
where  $v$ is the flow velocity and $c$ the speed of sound.
For stationary flows, $c$ and $v$ 
only depend on 
$x$.
Assuming that the 
condensate flows from right to left ($v<0$),  
a sonic 
horizon is present where $w = c+v$ crosses $0$.
It corresponds to a black hole horizon
when $\partial_x w > 0$, and a white one when $\partial_x w < 0$.
We shall call it 
a
Killing horizon since 
the norm of the Killing field $\partial_t$ vanishes when $w=0$~\cite{lr}. 
Its location 
is taken at $x=0$.

When ignoring short distance dispersion,
the spectrum of the upstream phonons 
spontaneously emitted from a black hole 
horizon is 
simple,
and strictly corresponds to the Hawking 
radiation~\cite{Unruh81,Unruh95}. 
It follows a Planck law and the temperature is fixed by the 
{decay rate}~\footnote{To adopt the standard 
terminology, we shall call $\kk$ the {\it surface gravity}~\cite{BCH},
even though the latter has a dimension of an acceleration, 
and has no clear physical meaning 
for
sound waves.}
\begin{equation}\label{eq:temprelnew}
\kk \equiv \partial_x (c+v) \vert_{x = 0}. 
\end{equation}
For relativistic fields indeed, 
$\kk$ 
gives the late time decay rate of the frequency of out‫going 
modes, 
as seen by asymptotic observers, 
see Eq.~(32.8) in~\cite{MisnerThorneWheeler} for classical waves, 
and 
Eq.~(3.44) in~\cite{Primer} for Hawking radiation. 
As a result, $\kk$ fixes the temperature
to be 
exactly $\Th 
= \kk/ 2 \pi $, 
in units where $\hbar = k_B = 1$, but
independently of 
$\chor$, the value of the sound speed at the horizon.

When including dispersion, 
these results are no longer true because 
the total amount of redshift saturates. 
Nevertheless, the 
$\kk$-decay law is recovered 
near the horizon 
where $w = c + v$ is linear in $x$. Moreover, 
when the healing length is much smaller than 
$c_0/\kk$, this 
 guarantees 
that 
the spectrum and the correlations of the 
Hawking 
pairs 
are not significantly affected by dispersion~\cite{BMPS95,Rivista,
From2010}.

As already mentioned, 
the difficulties start when one tries to 
characterize the deviations 
from the Planck spectrum
at 
temperature $\Th$. This is because these 
have 
different origin and are governed by different parameters. In spite of this, 
we shall see that they 
can 
be analyzed and understood.
To this end, 
we shall work with flow profiles 
governed by {\it several} 
scales, and 
of the form 
\begin{equation}\label{eq:velocity_dec}
 c(x)+v(x)= w(x) = \wB(x) + \wt(x),
\end{equation}
where $\wB$ 
is a reference background profile and $\wt$ 
a perturbation of smaller amplitude:
 $0 \leq | \wt(x)| \lesssim |\wB(x)|$.
The metric~\eqref{eq:metric} is then completely 
fixed by introducing $q$, 
\begin{eqnarray}\label{eq:cv}
 c(x)&=&\chor+(1-q) w(x),\nonumber \\
 v(x)&=&-\chor+q \, w(x) ,
\end{eqnarray}
which specifies how $c+v$ is shared between $c$ and $v$. 
In all simulations, 
we work with $q=1/2$,
because it minimizes the 
scattering between left and right moving phonons which is not related to the Hawking effect 
(see Fig. 11 in~\cite{MacherBEC}).
Hence working with $q= 1/2$ will ease the identification of the other, more intrinsic,
deviations with respect to the standard flux.
Similarly, to have well-defined asymptotic modes, we work with infinite condensates and 
with 
$c, \, v$ possessing 
well-defined 
values
for $x \to \pm \infty$.

\begin{figure}
 \hspace{2pt}\includegraphics{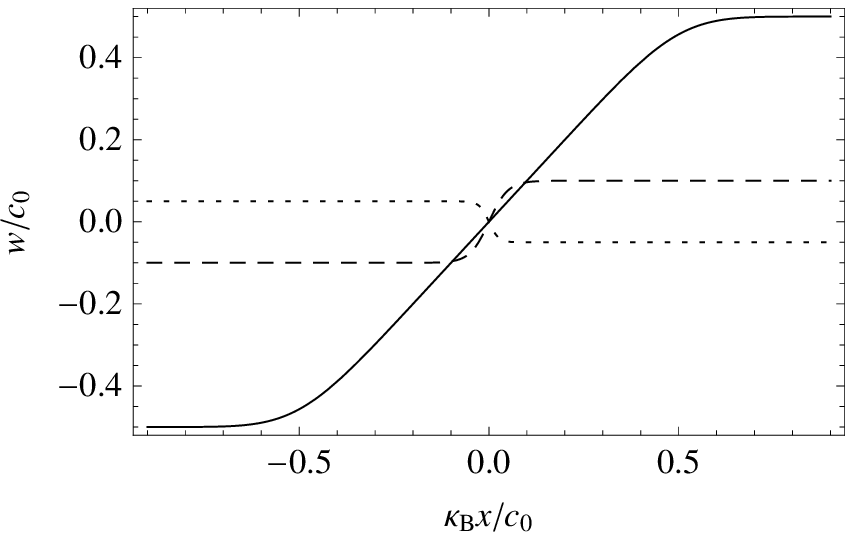}\\
 \hspace{6pt}\includegraphics{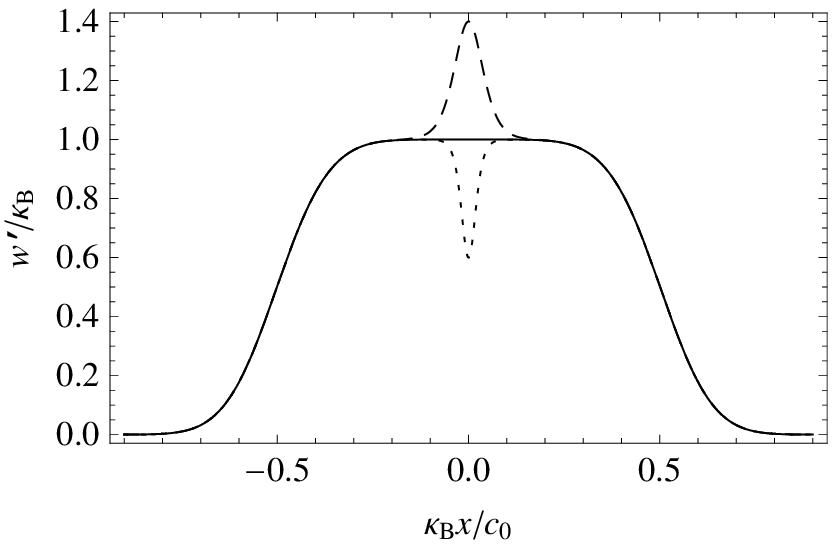}\\
 \includegraphics{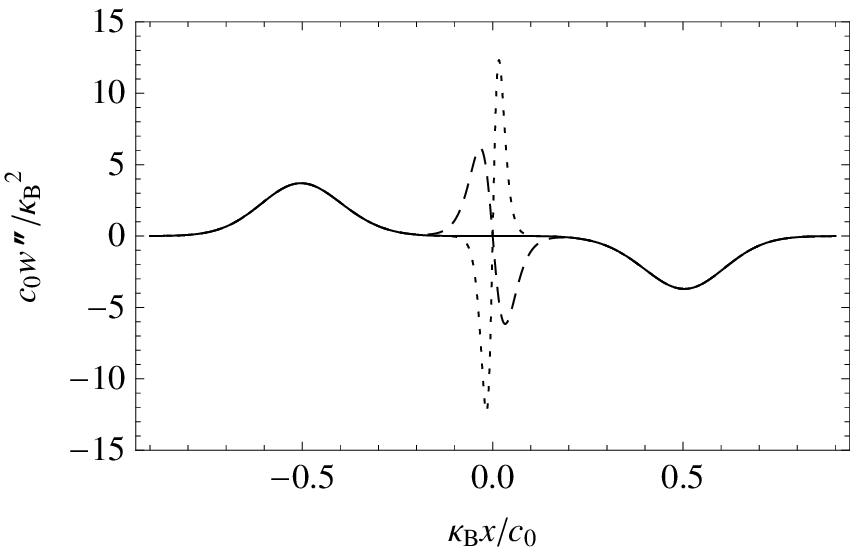}
 \caption{Upper panel: background flow 
$\wB$ (solid line) for $n=3$, $D=0.5$ and two 
perturbations $\wt$ (amplified by a factor of 5, dashed and dotted lines), with $\nt=1$,
and 
$(\kt/\ka, \Dt)$ equal 
to $(0.4, 0.02)$ and $(-0.4, 0.01)$. 
Central and lower panel: 
first and second 
derivative of $\wB$ (solid line) and of $w=\wB+\wt$ (dashed and dotted lines) for the same 
$\wt$. 
Narrower perturbations induce higher modifications of derivatives of $w$.
}
\label{fig:w}
\end{figure}

The background profile is given by
\begin{equation}\label{eq:velocityB}
 \frac{\wB(x)}{\chor} =  
D\times \sign(x) 
\left[ \tanh\left\{\left(\frac{\ka|x|}{\chor D}\right)^n\right\}\right]^{1/n}.
\end{equation}
The quantity $D$ fixes the asymptotic value of $\wB$, whereas $n$ governs the 
smoothness of the transition from the region where $\wB$ is linear in $x$
to the asymptotic regime where it is constant. We shall work with 
$n =1$ or $2$ 
to have a well-defined range $\sim \chor D/\ka$ 
where $\wB$ is linear, 
and yet avoid the nonadiabatic effects~\cite{CJ96,MacherRP1} 
when the transition is too sharp, 
i.e., when $n \geq 4$. 

We shall consider two 
types of perturbations,
symmetric ones with respect to the horizon, i.e., $\wt(x)= - \wt(-x)$,
and then, in Sec.~\ref{sec:assym}, 
more general ones without that symmetry. The symmetric ones 
are similar to 
$\wB$:
\begin{equation}\label{eq:velocity2}
\frac{\wt(x)}{\chor} = 
\Dt\times \sign(x) 
\left[ \tanh\left\{\left(\frac{\kt|x|}{\chor \Dt}\right)^{\nt}\right\}\right]^{1/{\nt}}.
\end{equation}
The condition $0 \leq | \wt| \lesssim |\wB|$ 
is simply 
$0 \leq \Dt \lesssim D$.
In Fig.~\ref{fig:w}, 
we plot 
$w(x)$, its first 
and second derivatives
 for different values of $\kt/\ka$ and $\Dt$. 

The smoothness parameter $\nt$ 
will range from $0.5 $ to $4$. 
We shall
see that it only 
affects the flux marginally. In fact the most important quantity is
$\Dt$.
Its role 
 is to fix 
\begin{equation}\label{eq:xlin}
|x_{\rm lin}| \sim  \frac{\chor \Dt}{\kt},
\end{equation}
the interval of $x$ over which $w(x)$ is
linear.
%
In this respect, 
it should be noticed that the 
surface gravity
is equal to
\begin{equation}\label{eq:temprel}
\kk \equiv \partial_x (c+v) \vert_{x = 0} = \ka + \kt,
\end{equation}
{\it irrespectively} of the value of $\Dt$, and that of $\nt$. 
For relativistic fields, 
$\Dt$ would play no role since 
the temperature is $\kk/2\pi$  
(when ignoring gray body factors, a correct approximation in our settings when $q=1/2$). 
For dispersive fields instead, 
$\Dt$ plays a crucial role.

\section{Spectral analysis}
\label{sec:numerics}

In a stationary 
condensate flowing with a velocity $v$,
the dispersion relation of phonons 
is given by~\cite{DalfovoRMP} 
\begin{equation}\label{eq:dispersion_hom}
(\om-vk)^2=\Omega^2(k)=
c^2 k^2 + \frac{\hbar^2 k^4}{4 m^2 } = c^2 k^2 +  \frac{c^4 k^4}{\Lambda^2},
\end{equation}
where $\om$ is the conserved 
frequency $\partial_t$,
$m$  the atom mass, and $\Lambda$ the frequency associated with the 
healing length $\xi = \hbar/\sqrt{2}mc$ by $\Lambda = \sqrt{2}c/\xi$. 

In what follows we study the properties 
of
$n_\omega$, the mean occupation number
of phonons of frequency $\omega$ 
spontaneously emitted 
to the right 
region, 
i.e. against the flow, when the initial state is vacuum. 
In order to focus on new results, 
we have decided to present in the Appendixes
a summary of the concepts needed to compute $n_\omega$, 
starting from the 
BdG equation applied to the flows of Eq.~\eqref{eq:velocity_dec}.
The figures of this paper 
have been 
obtained by numerically solving 
Eq.~\eqref{eq:system} with 
the code of Ref.~\cite{MacherBEC}. 

Our program can now be clearly defined: we shall 
study $n_\omega$ in the perturbed flows 
for increasing 
 $\Dt$ 
 to see how the phonon 
field
``responds'' to the introduction of 
 $\wt$.
We start with the 
symmetrical profiles of Eq.~\eqref{eq:velocity2}, and then consider asymmetrical ones.
The spectrum in the unperturbed flow of Eq.~\eqref{eq:velocityB} has been 
studied 
in~\cite{CJ96,MacherRP1,MacherBEC}.

\subsection{Planckian character}
\label{sec:general}
\begin{figure}
 \includegraphics{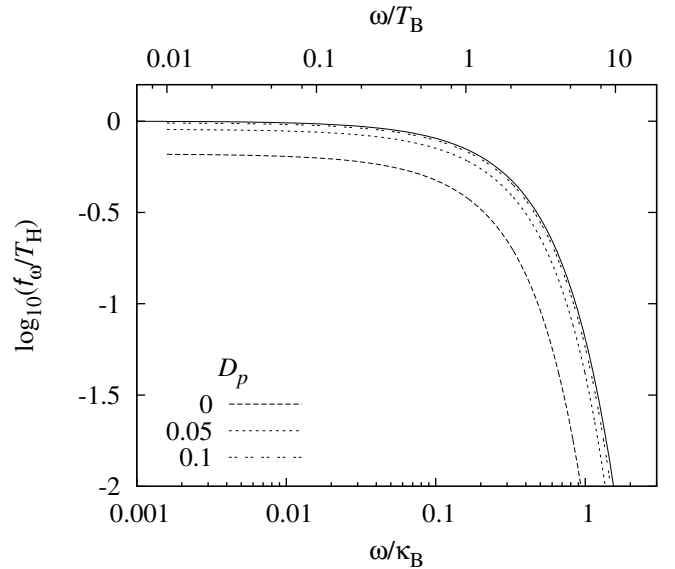}
\caption{$\log_{10}(f_\om/\Th)$ 
versus $\omok$ for different values of $\Dt$, 
but the same surface gravity $\kk$. 
The solid line represents the Planck flux at temperature $\Th=\kk/2\pi$. 
When $\Dt= 0.1$, the phonon flux hardly differs from it.
For smaller values of $\Dt$, the flux 
is weaker, and for $\Dt \to 0$, it is fixed by the background surface gravity $\ka$. 
The fixed 
parameters are $\Lambda/\ka = 15$, $D = 0.4$, $\kt/\ka=0.5$, $n=\nt=1$.
\label{fig:tom_fom_minus}
}
\end{figure}

In Fig.~\ref{fig:tom_fom_minus} we 
present 
the energy flux 
\begin{equation}
f_\om = \om \, n_\om , 
\end{equation}
as a function of $\om$ for different values of the 
amplitude  
$\Dt$, 
keeping fixed all the other parameters. 
%
When working with a relativistic massless field, 
one would obtain a 
Planck spectrum 
with a temperature 
$\Th = \kk/2\pi$, irrespectively of the value of $\Dt$. 
To ease the comparison with the dispersive fluxes,
we have represented this flux by a solid line. 
Working with 
the BdG 
equation~\eqref{eq:system}, 
we see instead that 
the flux varies with $\Dt$. 
In fact it  increases monotonically.  
Moreover, when $\Dt \to 0$, 
we verified that it 
coincides with the 
flux 
one 
 would obtain 
in the reference 
flow $\wB$.  
We also see that when $\Dt$ is sufficiently large, the flux 
saturates and agrees to a high accuracy
with the Planck spectrum at temperature $\Th$, 
in accordance with the 
{\it robustness}~\cite{Unruh95,BMPS95,CJ96} of
the flux.

Since comparing energy fluxes $f_\om$ is not 
convenient, 
we shall use 
the temperature function 
defined by
%
 \begin{equation}\label{eq:Tom}
  n_\om \equiv \frac{1}{\exp(
\om/ T_\om)-1}.
 \end{equation}
%
As such, $T_\om$ is simply another way to express 
$n_\om$.
However it presents the great advantage of being constant whenever the flux is Planckian.
\begin{figure}
  \includegraphics{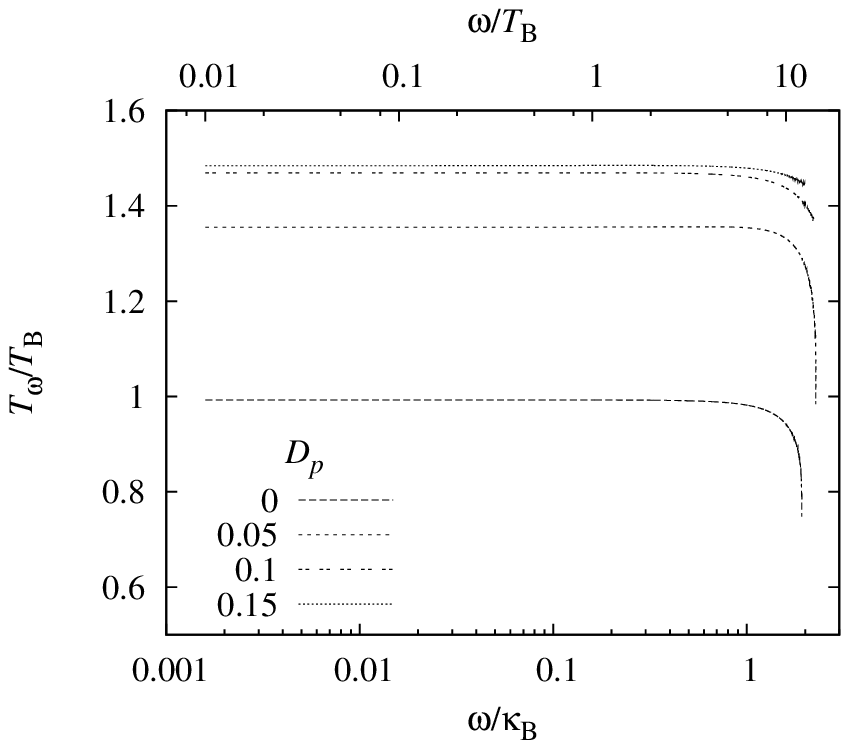}
   \caption{$T_\om/\TB$ of Eq.~\eqref{eq:Tom} versus $\omok$ for various 
$\Dt$ 
but the same value of $\kk= 1.5 \ka$.  
$\TB=\ka/2\pi$ is the Hawking temperature of the background flow. 
The values of the parameters are those of 
Fig.~\ref{fig:tom_fom_minus}. 
For all values of $\Dt$, $T_\om$ hardly varies 
until 
$\om \to \omm$ of Eq.~\eqref{eq:omm}, 
 where the flux vanishes. 
Since 
$\omm \gtrsim 10 \, \TB$, as can be seen from the horizontal upper scale, 
these spectra are, to a high precision, Planckian.
}
\label{fig:tom_fom}
\end{figure}

In fact,
as can be seen in 
Fig.~\ref{fig:tom_fom}, for {\it all} values of $\Dt$, 
$T_\om$ remains remarkably 
constant, until $\om$ approaches  
the critical frequency $\omm$ of Eq.~\eqref{eq:omm}. 
At that frequency, 
the signal drops down as it must do, since 
the Bogoliubov transformation \eqref{eq:bog_transf} becomes trivial,
because the negative norm mode $\varphi_{-\om}$ no longer exists above $\omm$.
This cutoff effect relies on the asymptotic value of the flow $v+c$, and
has been studied in detail in~\cite{MacherRP1,MacherBEC}. In what follows we focus instead
on the deviations of the spectrum which depend on the near horizon properties
of $v+c$. 


To quantify the small 
deviations  
from a Planck spectrum at temperature
\begin{equation}
  T_0\equiv\lim_{\om\to0}T_\om,
\end{equation}
we studied the relative deviation
 \begin{equation}\label{eq:run}
  \run\equiv \frac{T_0-T_{\om}}{T_0}.
 \end{equation}
evaluated for $\om = T_0$. 
For all values of $\Dt$ we found
\begin{equation}\label{eq:run2}
  |\runT| \lesssim 3.\times 10^{-4}.
 \end{equation}
We note that $\runT \neq 0$ is not 
due to numerical errors, it really characterizes the deviations. 

Equation~\eqref{eq:run2} 
implies that, 
to a high accuracy, 
the spectrum remains
Planckian
even when $T_0$ 
completely differs from $\Th$. (In the present case, it differs by $33 \%$, 
since $\kk = 1.5\, \ka$.)
The smallness of 
$\runT$
also implies that the parameters that govern 
the deviations from 
Planckianity {\it differ} from those governing the {shift} 
$T_0 
- \Th$. 
Hence when studying the robustness of the Hawking flux against introducing dispersion,
one must differentiate these two types of deviations. 

These are our first important results. 
They 
generalize what was observed in~\cite{MacherRP1,MacherBEC} to a much wider class of flows, 
namely, first, 
that the thermality is well preserved whenever 
 $\omm/\TB \gtrsim 10$,
and, second, that the relative 
temperature shift $T_0/\Th - 1$ is 
much larger than $\runT$. 
In the sequel we work with 
$\omm/\TB \gtrsim 10$
satisfied, 
and study 
how the 
 low frequency temperature 
$T_0$ 
depends on $\kt$, $\Dt, \nt$ and $\Lambda$. %
%

%
%

%


\subsection{The critical value 
of $\Dt$}
\label{sec:T0}

\begin{figure}
  \includegraphics{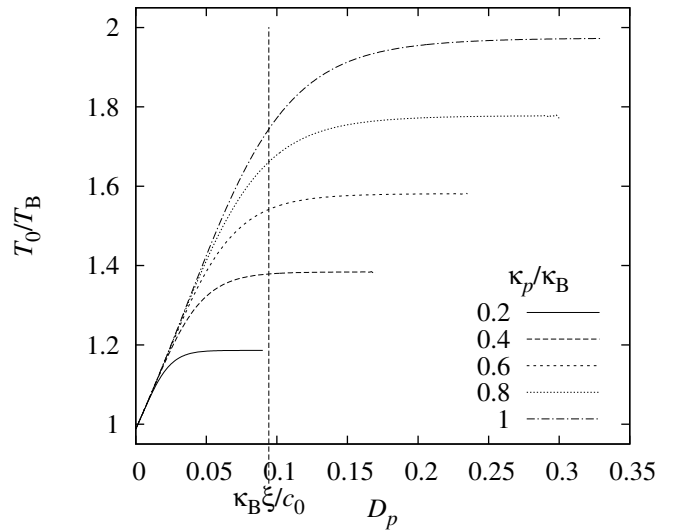}
   \caption{
$T_0/\TB$ versus $\Dt$  for 
$\kt/\ka$ from $0.2$ to $1$, and with $\Lambda/\ka = 15, D = 0.3$, $n=\nt=1$.
Along each curve the surface gravity $\kk$ of Eq.~\eqref{eq:temprel} is {\it constant}.
It fixes the temperature {\it only} for $D \gg \dcrit$.
The vertical line gives the value of the healing length $\xi$. }
   \label{fig:T0_k3}
\end{figure}

In Fig.~\ref{fig:tom_fom},
we saw that $T_0$ 
starts 
from $\ka/2\pi$ for $\Dt\to0$, 
 and asymptotes to $\Th = \kk/2\pi$ 
for $\Dt$ larger than some critical value $\dcrit$. 
(We have checked that this is also the case when $\kt < 0$, and this down
to $\kt = - \ka$, in which case the surface gravity $\kk$ vanishes.)
To understand 
what 
fixes $\dcrit$,
we first consider 
series of flows $\wB + \wt(\Dt)$ 
for different values 
of $\kt$ at fixed $\ka$. 
For each series, the surface gravity $\kk = \ka + \kt$ is thus constant.
The resulting temperatures $T_0(\Dt)$ 
are presented in Fig.~\ref{fig:T0_k3}. 
This figure 
reveals many interesting features.
%
For small values of $\Dt$, 
$T_0$  increases linearly with $\Dt$ 
in a manner essentially {\it independent} of 
$\kt$.  
In addition, for each series, when 
$\Dt$ is large enough, 
$T_0$ saturates 
at the standard result 
$\Th = \kk/2\pi$.

Given the linear behavior of $T_0$ for small $\Dt$ and the saturation for 
large values, i.e.,
\begin{equation}\label{eq:limits}
\begin{aligned}
2\pi T_0 
(\Dt\ll\dcrit)&\sim \ka+s\, \Dt, \\
2\pi T_0 
(\Dt\gg\dcrit)&\sim \ka+\kt=\kk,
\end{aligned}\end{equation}
%
we define $\dcrit$ as the value of $\Dt$ where these 
intercept,
and obtain
%
\begin{equation}\label{eq:dcrit}
 \dcrit = \frac{\kt}{s}.
\end{equation}
To understand what fixes the slope $s$,
%
\begin{figure}
  \includegraphics{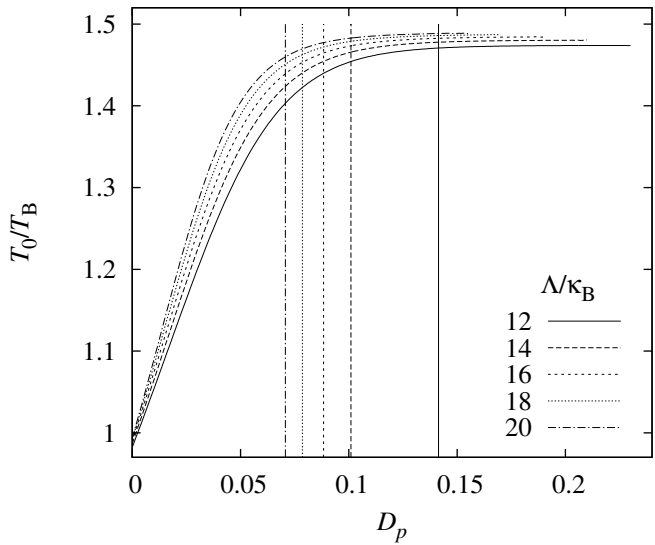}
   \caption{
$T_0/\TB$ versus $\Dt$ for 
$\Lambda/\ka$ from $12$ to $20$, and with $\kt/\ka = 0.5, D = 0.3$, $n=\nt=1$.
The vertical lines give the values of the healing length $\xi = \sqrt{2}\chor/\Lambda$.
One can see that 
$\dcrit$ of Eq.~\eqref{eq:dcrit1}
is {\it not} (directly) related to $\xi$.}
   \label{fig:T0_k3_bis}
\end{figure}
we consider series of flows 
with different values of  
the ultraviolet dispersion scale $\Lambda$.
The results are presented in Fig.~\ref{fig:T0_k3_bis}.
It is numerically easy to see 
that 
$s\propto \Lambda^{2/3}$. 
Hence we get 
%
\begin{equation}\label{eq:dcrit1}
 \dcrit\propto \kt\, \Lambda^{-2/3}.
\end{equation}
%

We now present a
simple 
argument telling 
how $\dcrit$ should 
scale with 
 the various parameters. 
In~\cite{MacherBEC}, by considering $\wB$ of Eq.~\eqref{eq:velocityB}, it was 
{\it observed}
that the leading deviations due to dispersion are governed by inverse powers of $\ommok$.
For these unperturbed flows, 
$\omm$ of Eq.~\eqref{eq:omm}, even though initially defined as the frequency such that 
the turning point $x_{\rm tp}(\omega)$ of Eq.~\eqref{eq:tp}
is rejected at $- \infty$, 
also 
corresponds to the frequency such that the turning point is at the edge of the 
domain $\sim \chor D/\ka$, 
where $\wB$ 
is linear in $x$. 

When working with 
$w= \wB + \wt$ of Eq.~\eqref{eq:velocity2} and with $\Dt \ll D$,
the linear domain is now given by $x_{\rm lin}$ of Eq.~\eqref{eq:xlin}.
Hence 
we expect that the deviations will be now governed by inverse powers of $\ommt/\kk$, 
where $\ommt$ is the new critical frequency fixed by $x_{\rm lin}$.
%
The value of $\ommt$ is such that 
the turning point $| x_{\rm tp}(\omega)|$ 
of Eq.~\eqref{eq:tp} equals $x_{\rm lin}$.
%
Using 
Eq.~\eqref{eq:xlin}, 
 $| x^{\rm tp}(\ommt)| = x_{\rm lin}$ 
 gives %
\begin{equation}\label{eq:omm2}
 \ommt =\Lambda\left( \frac{\kk}{\kt} \Dt\right)^{3/2}.
\end{equation}
Since the deviations from the standard result
should be small when $\ommt/\kappa_K\gg1$,
$\dcrit$ should scale as
\begin{equation}
\dcrit \propto \frac{\kt}{(\kk\, \Lambda^2 )^{1/3}}.\label{eq:dcrittheo}
\end{equation}
A rigorous study~\cite{ACRP2} of the connection 
formulas~\cite{BMPS95,Corley97,Tanaka99,UnruhSchu05}
 confirms 
the result of this simple reasoning. 


%
%
Equation~\eqref{eq:dcrittheo} is in agreement with 
Eq.~\eqref{eq:dcrit1}
for $\kt \ll \ka$. 
%
In the following section, we shall numerically validate it
for all values of $\kt$. 
Before proceeding, we checked that 
$\dcrit$ does not 
significantly depend on the 
background quantities $n$, $D$ and $\ka$, 
as one could have expected.
The weak, subleading, dependence on 
$\nt$ 
is studied in Sec.~\ref{sec:n2}.

\subsection{The averaged surface gravity}
\label{sec:bark}

So far we have some understanding 
of the Planckian character of the flux 
in the robust regime when $\Dt > \dcrit$, and of the 
value of $\dcrit$~\cite{ACRP2}. 
Instead, what fixes 
$T_\om$ 
outside this regime, for $\Dt < \dcrit$, is {\it terra incognita}.

We shall now numerically establish that, to a 
good approximation when $\omm/\TB \gtrsim 10$, 
the temperature is determined by the average of
the gradient $dw/dx$ over a 
width 
that we parametrize by $d_{\xi}=\dxl + \dxr $,
where $\dxl$ and $\dxr $ are respectively the width on the left and on the right calculated
from the Killing horizon at $x=0$.
 
In fact, given that the dispersion of Eq.~\eqref{eq:dispersion_hom} 
defines a `healing length' $\xi = \sqrt{2}c/\Lambda$, 
it is not 
surprising that $\xi$ 
 in turns defines 
a minimal resolution length, such that details of $w(x)$  
smaller than $d_{\xi}$ 
are not ``seen'' by the phonon field.
In other words, because of dispersion, it is as if we were dealing with a 
horizon
of width $d_{\xi}$.\footnote{\label{foot:hor}This important result
should be opposed to the possibility discussed in~\cite{UnruhSchu05,08}
according to which $T_\om$ could be governed by a local, $\om$-dependent, function of $w$, e.g. by the
value of  $\partial_x w$ evaluated at the turning point of Eq.~\eqref{eq:tp}.  
In fact our numerical results indicate 
that it is a {\it non-local} quantity that fixes the temperature, 
and in a manner essentially independent of $\om$. 
This 
is in sympathy with the ideas that quantum gravitational effects might blur the horizon~\cite{quantum_metric_fluctRP,Beyond_RP} and that
dispersion might regulate the black hole entanglement entropy~\cite{ent_ent_TJRP}.}
Moreover, when $\dxl \neq \dxr $, this means that the center of the effective horizon is {\it displaced} 
with respect to the Killing horizon to $\bar x_{\rm hor}= (\dxr - \dxl)/2$.

%
To test the idea that the temperature be determined by an average ``surface gravity,'' we introduce 
\begin{eqnarray}\label{eq:kav}
 \bar\kappa & \equiv & \frac{1}{d_{\xi}} \int_{-\dxl}^{\dxr} \dd x \, \frac{\dd w(x)}{\dd x}= 
 \frac{w(\dxr) -w ( -\dxl) }{d_{\xi}} .
\end{eqnarray}
Using 
Eqs.~\eqref{eq:velocityB} and~\eqref{eq:velocity2}, we get
\begin{eqnarray}\label{eq:kavs}
 \bar\kappa
&=& \ka+ \frac{\wt(\dxr)  + \wt(\dxl) 
}{d_{\xi}} 
+O\left[\left(\frac{\ka d_{\xi}}{Dc_0}\right)^{2n}\right],\quad 
\end{eqnarray} 
where we assume that $d_\xi \ll D/\ka$, so that the average of the background term $\wB$ 
is approximately $\ka$. 

\begin{figure}
 \includegraphics{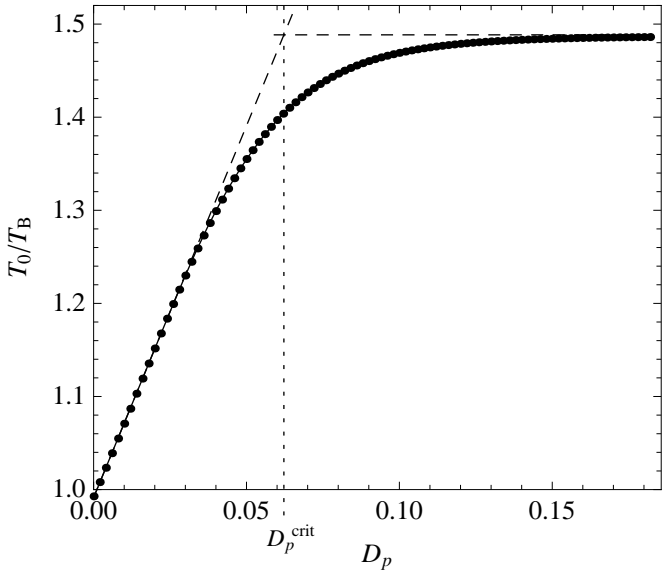}
 \caption{Dots: $T_0/\TB$ versus $\Dt$ computed 
for $\Lambda/\ka=15$, $D=0.4$, $\nt=n=1$, $\kt/\ka=0.5$.
Solind line: best fit curve obtained by using 
$\bar \kappa/2\pi$
of
Eq.~\eqref{eq:kavs}. 
The perfect agreement 
establishes that the 
temperature $T_0$ equals $\bar \kappa/2\pi$.
Dashed lines: asymptotic behaviors for $\Dt\ll\dcrit$ and $\Dt\gg\dcrit$ [see Eqs.~\eqref{eq:limits} and~\eqref{eq:dcrit}].}
 \label{fig:fitbark}
\end{figure}

We proceed as follows: 
We compute $T_0$ for about 200 values of $\Dt$
for some fixed values of the 
perturbed flow parameters 
$(\nt, \kt)$ and the dispersive scale $\Lambda$, i.e., along  series 
as those represented in Figs.~\ref{fig:T0_k3} and~\ref{fig:T0_k3_bis}. 
Out of this set 
we extract five quantities, namely, $\ka^{(\rm fit)}$, $\nt^{(\rm fit)}$, $\kt^{(\rm fit)}$, $\dxl$, and $\dxr$, 
by fitting 
$\bar \kappa(\ka,\nt,\kt,\dxl, \dxr;\Dt)/2\pi$ of Eq.~\eqref{eq:kavs}
with a nonlinear least-squares method.
The agreement between the 
couples $T_0(\Dt)$ 
and 
the 
fitted function 
$\bar \kappa(\Dt)/2\pi$ 
is really striking---see Fig.~\ref{fig:fitbark} for an example.
\begin{figure*}
  \includegraphics{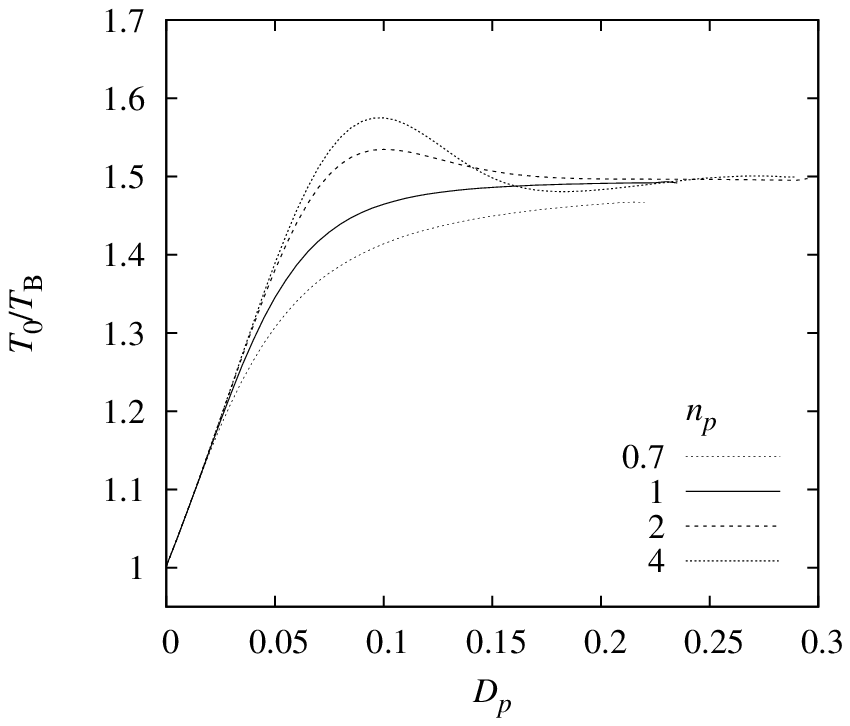}
  \hspace{10pt}
  \includegraphics{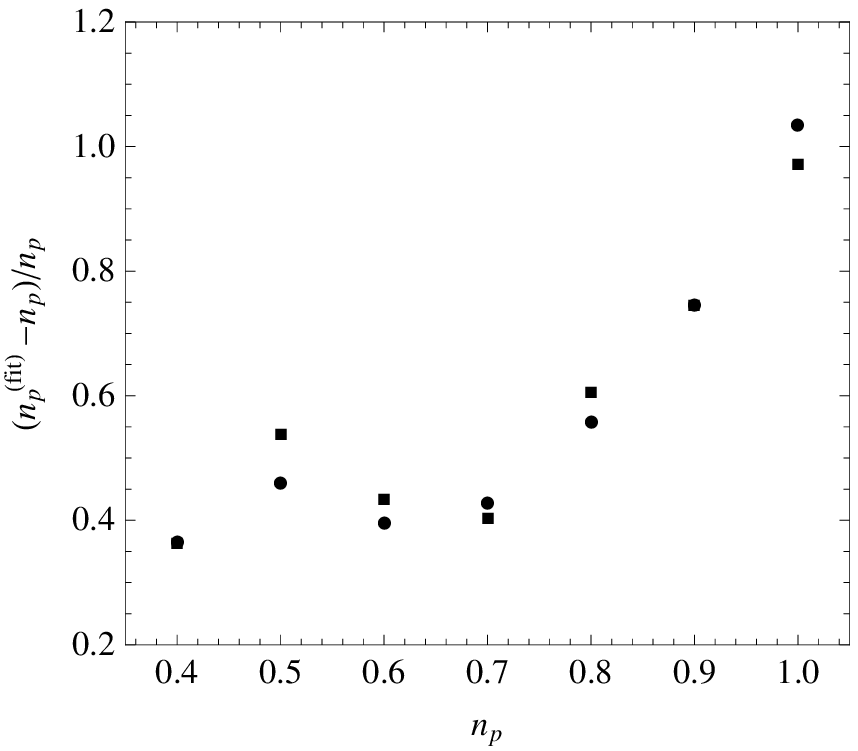}
   \caption{Left panel: $T_0/\TB$ versus $\Dt$ for different values of $\nt$ of  Eq.~\eqref{eq:velocity2} with 
$\kt/\ka = 0.5$, $D = 0.4$, $n=2$. 
$T_0$ monotonically approaches 
$\Th$ for $\nt\leq1$. When $\nt>1$ instead, it exceeds $\Th$ near $\dcrit$ and then approaches it 
with oscillations.  Right panel: 
the relative difference $(\nt^{\rm (fit)}-\nt)/\nt$ versus $\nt$. The difference 
decreases for smaller $\nt$, 
in a rather insensitive way with respect to $\Lambda/\ka$ (dots, $\Lambda/\ka=15$; squares, $\Lambda/\ka=20$). }
   \label{fig:n2}
\end{figure*}

Moreover, 
the fitted values $\ka^{(\rm fit)}$ and $\kt^{(\rm fit)}$ 
are in very good agreement (less than 1.5\% and 1\%, respectively) with their values used in the flow $w$.
This is a necessary check to validate that 
it is the average of $\partial_x w$ 
 that governs the temperature.
However, 
the agreement between 
$\nt^{(\rm fit)}$ and $\nt$
is less good. This indicates that the actual 
temperature is not 
{\it exactly} given by 
the average of Eq.~\eqref{eq:kav}.~\footnote{We notice that 
different averaging procedures could have been considered in the place of Eq.~\eqref{eq:kav}, and they
would have given slightly different results. 
At this point, we are not in the position to distinguish them.
What is important is 
that the scaling laws we shall derive in Sec.~\ref{sec:dxi} are independent 
of the particular choice.} 
Even though this is
a 
subdominant effect, 
it is interesting to see under which conditions
a better agreement between $\nt^{(\rm fit)}$ and $\nt$ is reached.

\subsection{Subleading effects in $\nt$}
\label{sec:n2}

The parameter $\nt$ of Eq.~\eqref{eq:velocity2} determines the smoothness of the 
transition between the region 
where 
$\wt(x)$ is linear, with constant derivative $\kt$, 
and the region where it is flat, equal to $\Dt c_0$. Larger values of $\nt$ make that transition 
sharper. 
Therefore 
higher derivatives of $w$ will become very large at the end of the linear-regime region, near 
$x_{\rm lin}$ of Eq.~\eqref{eq:xlin}. 
As a consequence, nonadiabatic effects become more important, 
thereby generating larger
deviations between 
$T_0$ and $\bar \kappa/2\pi $
of Eq.~\eqref{eq:kavs}.



To investigate these deviations, 
we compare $T_0(\Dt)$ for various values of $\nt$. 
As it appears from Fig.~\ref{fig:n2}, left panel, $s$ and $\dcrit$ do not significantly depend on $\nt$. 
However, the transition region between the 
linear regime and the 
constant regime 
depends on 
$\nt$. The transition is indeed monotonic for small values of $\nt$
whereas, 
for $\nt > 1$, 
  $T_0$ 
oscillates around 
$\Th = \kk/2 \pi$.  
The monotonic function of Eq.~\eqref{eq:kav} will of course never 
describe these oscillations.
However, 
for $\nt\leq1$, one can 
compare $\nt^{(\rm fit)}$ with $\nt$. 
The result is shown in Fig.~\ref{fig:n2}, right panel. The dots and the squares are computed using 
different values of $\Lambda/\ka$, respectively 15 and 20. 
One sees that $\nt^{\rm fit}$ 
is always 
larger than 
$\nt$. 
One also sees that 
the discrepancy becomes smaller 
when $\nt$ decreases, as expected,  
since nonadiabatic effects are smaller for smoother profiles, hence smaller $\nt$. 
One should not overestimate the relevance of this discrepancy because we found that the 
value of $\nt^{(\rm fit)}$ is correlated in the fitting procedure to the difference $\dxr - \dxl$.
For instance, when imposing $\dxr = \dxl$, $\nt^{(\rm fit)} - \nt$ is reduced by a factor of 
about 
$2$.


The important conclusion 
is the following.
When nonadiabatic effects are 
small,
the temperature of the phonon flux is well approximated 
by $1/2\pi$ times the average of $\partial_x w$
over a certain width $d_\xi$ [see Eq.~\eqref{eq:kav}].

\subsection{The scaling laws 
of the width $d_\xi$}
\label{sec:dxi}

Unlike $\ka$, $\kt, \nt, \Dt$ which characterize the flow $w$
and $\Lambda$ which fixes the healing length, $d_{\xi}$ is not a 
parameter entering the mode equation~\eqref{eq:system}.
Therefore the value obtained in a fit cannot
be compared with some a priori known value.
In fact $d_\xi$ should be conceived as an {\it  emergent}, effective, length scale.

Before using numerical results to 
determine how it scales when changing $\kk$, $\kt$ and $\Lambda$,
we point out that we can predict how it should do,
as $d_\xi$ is closely related to $\dcrit$ of Eq.~\eqref{eq:dcrittheo}. 
Indeed, taking the limits of Eq.~\eqref{eq:kavs} for small and large $\Dt$ we obtain
\begin{equation}\begin{aligned}
 \bar\kappa(\Dt\ll 
\frac{\kt d_\xi}{\chor}
)&\sim\ka+\frac{2\Dt c_0}{d_\xi},\\
 \bar\kappa(\Dt\gg 
\frac{\kt d_\xi}{\chor})&\sim 
\kk,
\end{aligned}\end{equation}
and hence, using Eq.~\eqref{eq:limits} and~\eqref{eq:dcrit}, we get 
\begin{equation}\label{eq:hdcrit}
 d_\xi = \frac{2c_0\dcrit}{\kt}.
\end{equation}
%
In other words, the critical value of $\Dt$ separating the two regimes of $T_0(\Dt)$ shown in Fig.~\ref{fig:fitbark}
strictly corresponds to the width $d_\xi$ one should use in Eq.~\eqref{eq:kav} to follow $T_0(\Dt)$.
Using 
Eq.~\eqref{eq:dcrittheo}, we get 
%
\begin{equation}\label{eq:h}
 d_\xi = d \times \frac{c_0}{(\kk \, \Lambda^2)^{1/3}} ,
\end{equation}
where $d$ is a constant. In terms of the healing length $\xi = \sqrt{2}c_0/\Lambda$, this gives
\begin{equation}\label{eq:hh}
 \frac{d_\xi}{\xi} = d\times\left(\frac{2\xi}{d_K}\right)^{-1/3},
\end{equation}
where $d_K\equiv c_0/\kk$ is the surface gravity length.
Hence, when $\xi \ll d_K$,
$d_\xi$ is significantly larger than $\xi$.

The prediction 
of Eq.~\eqref{eq:h}
can be numerically validated 
by a {linear}~\footnote{When the function is linear in the parameters, a close-form solution of the 
$\chi^2$
 minimization problem is available. This method is therefore preferable to 
non-linear ones.} 
least-square fit of 
%
\begin{multline}\label{eq:hfit}
\log(d_\xi \frac{\ka}{c_0})
 =\log(d)+a\log\!\left(\frac{\Lambda}{\ka}\right)+b\log\!\left(\frac{\kt}{\ka}\right)
 \\+c\log\!\left(1+\frac{\kt}{\ka}\right).
\end{multline}
%
We used a grid of 110 values of $(\kt,\Lambda)$. 
The corresponding values of $d_\xi(\kt,\Lambda)$
are 
extracted from series of about 200 couples $(T_0,\Dt)$ using
the 
procedure described after Eq.~\eqref{eq:kav}. 
$\kt/\ka$ ranges from 0.1 to 1, with step 0.1 and $\Lambda/\ka$ ranges from 10 to 20 with unitary step. The other parameters are 
$D=0.4$, $n=\nt=1$, $q=0.5$.\footnote{Because 
of numerical noise, for some series 
($\Lambda$,$\kt$), it was not possible to obtain a sufficiently large number of points $(T_0,\Dt)$. 
 In particular it is difficult to obtain $T_0$ for large values of $\Dt$.
We therefore decided to use 
only the 
series 
for which we 
obtain $T_0$
for every value of $\Dt$ from $0$ to $0.1$
(namely 100 points, with step 0.001).
%
In this way, 
there are points with
$\Dt\gtrsim\dcrit$ in the series $(T_0,\Dt)$ used to determine $d_{\xi}$. 
This condition is satisfied by 71 series 
over 110.} 

The result of the fit is
\begin{equation}
\label{eq:hfitparams}
\begin{aligned}
 &a=-0.627,\quad &b=0.024,\\
 &c=-0.33,\quad &d=1.58,
 \end{aligned}
\end{equation}
%
where the numerical error is on the last figure. The exponent $a$ differs from about $5\%$ from its expected value $-0.667$. The exponent $b$ is also close to its expected value 0. Finally,
$c$ perfectly matches its theoretical estimate. 
Furthermore, we notice that the unknown constant $d$ introduced in Eq.~\eqref{eq:h} is of the order of 1.

In brief, 
for the considered class of flows, we have established that,
to some 
accuracy, 
\begin{itemize}
\item
the spectrum is Planckian, 

\item
the temperature equals 
$\bar \kappa/2\pi$ 
for a 
width $d_\xi$, and 

\item 
$d_\xi$ scales according to 
Eq.~\eqref{eq:h} with $d \simeq 1.6$.
\end{itemize}
%
%
These are the main results of this paper. 

From our fits we also found that the broadened horizon is slightly displaced 
with respect to the Killing horizon by a shift 
$|\bar{x}_{\rm hor}|\approx0.1d_\xi$.
However, because of the symmetry of Eq.~\eqref{eq:kavs}, we cannot know the sign of $\bar{x}_{\rm hor}$.
 To determine it, 
it is necessary to 
consider asymmetric flows. 


%


\section{Asymmetric flow profiles}
\label{sec:assym}

\subsection{Monotonic perturbations}

%
In this section, we consider 
profiles  of the form
\begin{equation}\label{eq:velocity2gen}
\wtl(x) = \wt(x-\delta) - \wt(-\delta),
\end{equation}
which are simply $\wt$ of Eq.~\eqref{eq:velocity2}
shifted from the horizon by $\delta$. 
We have subtracted the constant term 
so that $w_B + \wtl$ still vanishes at $x=0$.
\begin{figure*}
  \includegraphics{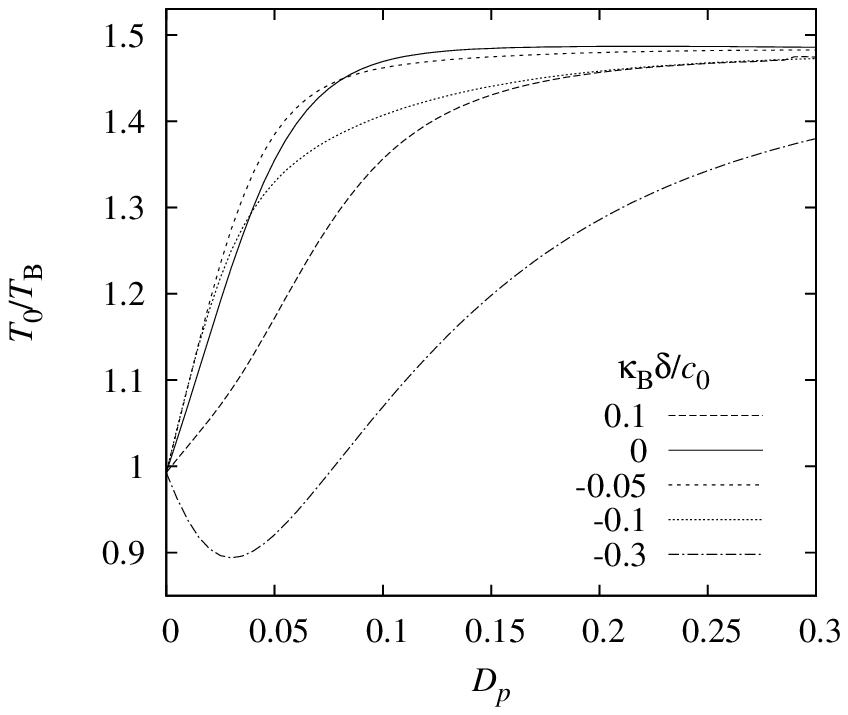}  
  \hspace{10pt}
\includegraphics{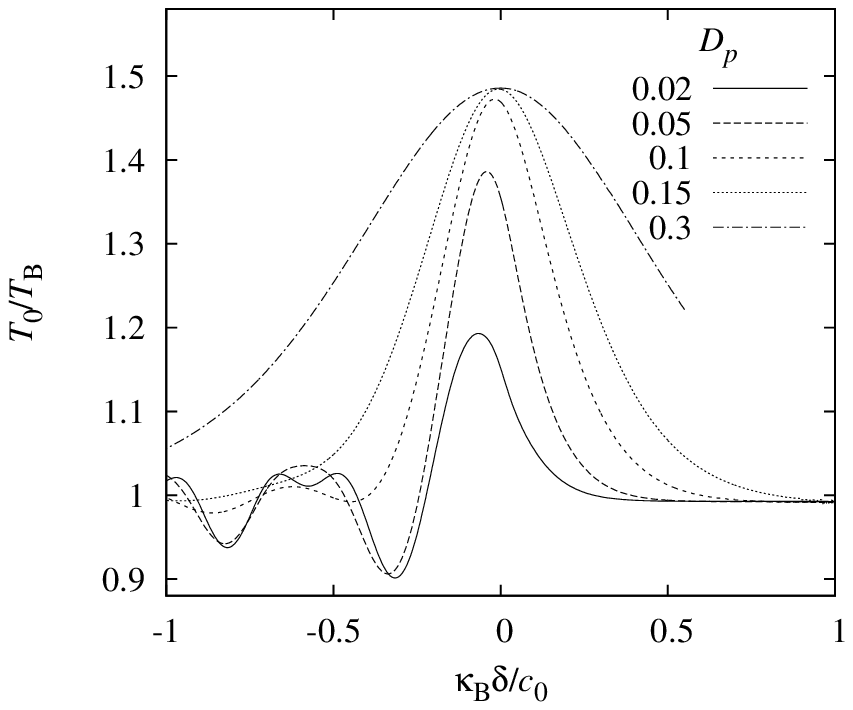}
   \caption{On the left panel,  $T_0/\TB$ as a function of $\Dt$  for various values of 
$\delta$, 
and, on the right, $T_0/\TB$ as a function of $\delta$ with 
 $\Dt$ from $\Dt=0.02$ to $\Dt=0.3$. $\Lambda/\ka = 15$, $D = 0.4$, $\kt/\ka=0.5$, $n=\nt=1$.
On the left panel, for values of $\delta \geq -0.05$, $T_0(\Dt)$ behaves as in Fig.~\ref{fig:fitbark}, 
whereas, for more negative values of $\delta$, $T_0$ behaves differently.
On the right panel, the asymmetry of $T_0$ under $\delta \to - \delta$ is clearly visible for 
$\Dt < \dcrit\approx0.06$. } 
  \label{fig:LS}
\end{figure*}
In Fig.~\ref{fig:LS}, 
$T_0$ is plotted as a function of $\Dt$ for different 
$\delta$ (left plot), and as a function of $\delta$ for different 
$\Dt$ (right plot). 
The behavior of these curves is quite complicated. To understand it, we first consider {\it smooth}
perturbations, i.e. perturbations with $\Dt>\dcrit\approx0.06$.

From the right panel of Fig.~\ref{fig:LS}, when $\Dt>\dcrit$, 
but for both signs of $\delta$, we see that $T_0$ monotonically interpolates 
from $\Th=\kk/2\pi$, when $\delta$ is small and the perturbation is close to the horizon, 
to $\TB=\ka/2\pi$, 
when $\delta\to\infty$. In that case, the perturbation is far from the horizon and thus no longer contributes to
$\kk$
 which equals its background value $\ka$.
When considering instead 
well localized perturbations, that is values of $\Dt$ smaller than $\dcrit$, 
we see 
that $T_0$ oscillates,
but only for $\delta$ sufficiently negative, i.e. when the perturbation is sufficiently displaced in the region
where the {\it decaying mode} 
oscillates.~\footnote{
We have adopted this formulation 
because, for subluminal dispersion, we expect that the sign of $\bar x_{\rm hor}$ will be the opposite,
and this in virtue of the `symmetry' between the behavior of modes for sub and super-luminal dispersion~\cite{ACRP2}.
Indeed, for (sub) super-luminal dispersion the turning point is in the (sub) super-sonic region, see Eq.~\eqref{eq:tp}.
In both cases, the mode decays on the other side of the turning point~\cite{CJ96,MacherRP1}.}
The origin of these oscillations can be understood as follows. 
The scattering on the well localized bump in $\partial_x w(x)$ interferes 
with the scattering on the horizon, thereby producing oscillatory behaviors.
\footnote{This is very similar to what has been found in inflation. When introducing a
sharp modification 
of mode propagation before horizon exit, there are interferences
between this localized scattering and the standard mode amplification which produce
oscillations in the power spectrum, see~\cite{CNP} for references and a critical analysis
on the generic character of this.}


It is quite clear that the average of Eq.~\eqref{eq:kav} cannot describe these oscillations.
However this should {\it not} be conceived as limiting the validity of the observation 
that the temperature is given by this average. Indeed, 
these oscillations result from interferences between two spatially 
separated scatterings. 
In these circumstances
it is meaningless to apply Eq.~\eqref{eq:kav}, which is based on the near horizon properties of $w$. 

In addition, 
when oscillations are found for $T_0$ as a function of $\Dt$ or $\delta$, 
the 
Planck character of the 
spectrum is {\it also} lost.
\begin{figure}
\includegraphics{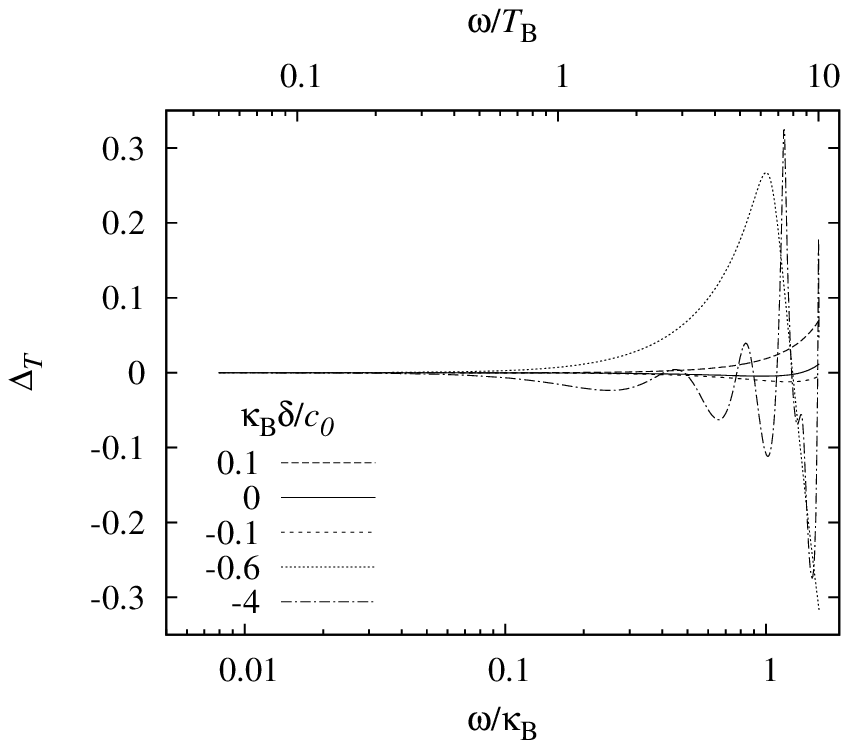}
\caption{The 
parameter $\run$ of Eq.~\eqref{eq:run} versus $\omok$ for different values of $\delta$. 
For $\delta=0$, continuous line, the perturbation is symmetric with respect to the horizon, and 
$\run$
is extremely small 
in agreement with Eq.~\eqref{eq:run2}. 
When $\delta > 0$ and $(-\delta) \leq d_\xi \approx 0.2 \, c_0/\ka $, 
$\run$
stays very small
 and no oscillations are found.
Instead when $-\delta \gg d_\xi$, 
oscillations of large amplitudes 
appear even for low frequencies, i.e. 
$\om \ll \omm \simeq 13 \,\TB$, thereby indicating that the spectrum is no longer Planckian.
The fixed parameters are $\Lambda/\ka = 15$, $D = 0.4$, $\Dt=0.02$, $\kt/\ka=0.5$, $n=\nt=1$. 
}
\label{fig:run_LS}
\end{figure}
Indeed, the parameter $\run$ of Eq.~\eqref{eq:run}
characterizing deviations from Planckianity also displays
 oscillations 
(of relative amplitude $\sim 20 \%$) (see Fig.~\ref{fig:run_LS}).
As can be seen, the number of peaks increases with increasing values of $-\delta$.
This is to be expected since more ``room'' is available to fit resonating modes. In fact this phenomenon 
can be viewed as a precursor of the black hole laser effect~\cite{CJlaser}. 
If the perturbation were strong enough to behave as a white horizon, 
the resonating modes would start growing exponentially~\cite{AC,bhlasers}. In that case
complex-frequency modes appear 
when the distance between the horizons is sufficiently large, and their number increases with that distance.

\begin{figure} 
\includegraphics{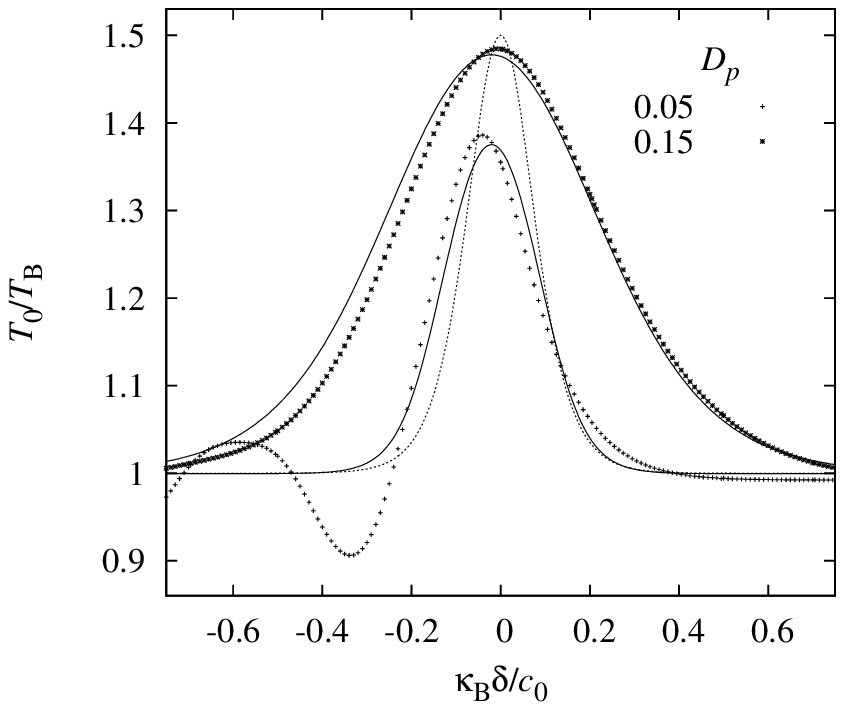}
   \caption{Numerical data (dots) for the series $T_0(\delta)/\TB$ 
with $\Dt=0.15\approx 2.5\dcrit$ and $0.05\approx0.8\dcrit$ 
   of Fig.~\ref{fig:LS}, right plot. The solid lines are  
the values of the fitted function of Eq.~\eqref{eq:kav}. 
The dotted line represents the surface gravity $\kk(\delta)/2\pi$ for $\Dt=0.05$.
The agreement between the temperature and 
$\bar \kappa/2\pi$
is quite good in the domains 
with no interference. 
The lowering of the peak and its shift to the left for small values of $\Dt$ are rather well 
reproduced. 
None of these effects are found when using the surface gravity $\kk(\delta)$.
} 
   \label{fig:LS_fit}
\end{figure}

To complete this analysis of the flows of Eq.~\eqref{eq:velocity2gen}, we verified that $\bar \kappa$ of Eq.~\eqref{eq:kav} 
still approximately governs the temperature. 
In Fig.~\ref{fig:LS_fit}, both the numerical data (dots) and the fitted function (solid line) are reported for the
series $T_0(\delta)$, with $\Dt=0.05$ and $0.15$, of Fig.~\ref{fig:LS} (right panel). This function is obtained as a two-variable function of $\delta$ and $\Dt$ by using the numerical data of $T_0$. The fitting region is restricted to those values of $\delta$ for which there are no oscillations and for three values of $\Dt$, namely 
$\Dt=0.05,0.1,0.15$. 
As a reference, the dotted line represents $\kk(\delta)/2\pi$, i.e. the temperature one would obtain without dispersion, for $\Dt=0.05$.
As a check, we also fitted  the values of $\ka$, $\kt$ and $\nt$ and compared them with the corresponding constants used in the numerical analysis. The relative difference is very small, respectively, around $0.1\%$, $1\%$ and $5\%$.

As one can see in Fig.~\ref{fig:LS_fit}, the agreement between the actual temperature $T_0$ and the average surface gravity $\bar \kappa/2\pi$ is quite good in the parameter region where there are no interference effects. 
One can also see that for $\Dt = 0.05$, $\bar \kappa/2\pi$ of Eq.~\eqref{eq:kav} follows $T_0$ much more closely than the 
surface gravity $\kk(\delta)/2\pi$. 



In summary, the analysis of asymmetrical 
perturbations first indicates
that the 
shift $\bar x_{\rm hor}$ 
is always negative. That is, the broadened horizon is displaced towards the region where the {decaying mode} 
oscillates. Moreover we find that the value of the shift does not depend directly on $\delta$. 
By this we mean that $\bar x_{\rm hor}$ depends on $\delta$ only because 
$d_\xi$ of Eq.~\eqref{eq:h} is a function of $\delta$ through $\kk(\delta)$.
Second,
for sufficiently smooth
perturbations we found that the temperature is still approximately given by the average of Eq.~\eqref{eq:kav}. 
Finally, 
when the perturbation is sharp and 
sufficiently deep inside the supersonic region, we found that the spectrum is affected by interferences,
and explained why it is so.

%





\subsection{Undulations} 
\label{sec:oscillation}


Supersonic flows possessing a sonic horizon with a negative {surface gravity} $\kk$
behave like the time reverse of black hole horizons: long wavelength modes are blueshifted
when scattered on such a horizon. Thanks to dispersion, the blueshift effect saturates,
and as a result, the scattering matrix is well-defined. Moreover, for each $\om$, 
this matrix is closely related to Eq.~\eqref{eq:bog_transf} evaluated
in the corresponding black hole flow~\cite{MacherRP1,MacherBEC}. 
In spite of this correspondence, it has been recently observed
that white hole flows display a specific phenomenon related to the limit $\om \to 0$~\cite{Carusottowhite,Silke}. 
Indeed, these flows produce a zero-frequency mode, a stationary undulation, 
which has a finite wavelength $k_Z$ and a high amplitude. 

To schematically describe such undulation and study its impact on the emitted spectrum,
we consider the following 
perturbations: 
%
\begin{equation}\label{eq:periodic}
\frac{\wt(x)}{\chor} = 
A\times  \sin\left(k_Z x-\theta\right) \times \ee^{-(x/\Delta)^4}.
\end{equation}
$A$ is the amplitude of the undulation, and $\theta$ fixes its phase on the 
horizon
at $x=0$. The new length $\Delta$ is taken to be larger than $1/k_Z$, and comparable to $D \chor/\ka$
which characterizes the linear regime of the background flow $\wB$.

\begin{figure}
  \includegraphics{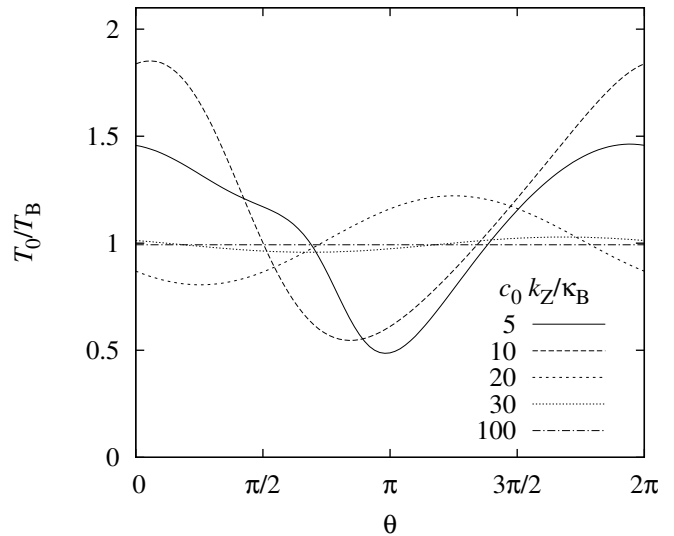}
   \caption{$T_0/\TB$ as a function of $\theta$ for 
Eq.~\eqref{eq:periodic}, 
varying $k_Z$ and $A$ so that 
$A\, k_Z=0.5\ka$ stays fixed. $\Delta=2 D \chor/\ka$, $\Lambda/\ka = 15$, $D = 0.4$, $n= 
1$.
For very high values of $k_Z$, i.e.  $k_Z = 100$ and $30$ $\ka/\chor$, 
the temperature is approximately $\ka/2\pi$, i.e. equal to $\bar \kappa/2\pi$ 
of Eq.~\eqref{eq:kav}.
For small values of $k_Z$, i.e. $k_Z =  5 \, \ka/\chor$, $T_0$ 
also approximately
follows $\bar \kappa(\theta)/2\pi 
\sim \kk(\theta)/2\pi$  
since the undulation wavelength is larger than $d_\xi$. 
In the intermediate regime instead, 
for $k_Z = 10$ and $20 \, \ka/\chor$,
the temperature  widely oscillates and no longer follows $\bar \kappa/2\pi $.
}
  \label{fig:T0_delta}
\end{figure}

In Fig.~\ref{fig:T0_delta}, $T_0$ is plotted versus $\theta$ for different values of $k_Z$,
keeping fixed $A\, k_Z$ so that 
the surface gravity of Eq.~\eqref{eq:temprel} does not change with $k_Z$. Indeed one gets
$\kk(\theta) = \ka+ A\,k_Z \cos(\theta) $.
%
When $k_z \gg 1/d_\xi$, as expected from our previous results, 
the temperature is very close to $\ka/2\pi$, that of the background flow $\wB$.
Hence, independently of the value of $\theta$, the perturbation is completely washed out
by the dispersion. 
On the contrary, when the wavelength is larger than the width of the horizon, 
i.e. when $k_z < 2\pi/d_\xi \approx 30 \ka/\chor$, 
we expect that 
$2\pi\, T_0$ will be approximately given by $\kk(\theta)$ given above. 
This can be 
observed for $k_Z$ equal to 
$5\, \ka/\chor$. 
One notices that 
 the curve is not exactly symmetric around $\theta=\pi$. 
But this is also expected from the former analysis
where we saw that moving the perturbation in the supersonic or in the subsonic region affects 
differently the spectrum.

The novelties are found for intermediate values of $k_Z$ 
(say, between $10\ka/\chor$ and $20\ka/\chor$). For these values, 
$\delta_Z = 2\pi/k_Z$ is comparable to 
$d_\xi$. Hence the undulation can enter into resonance with the scattering on the horizon,
thereby producing a kind of parametric amplification of the flux. 
This can be observed for
$\theta = 0 $ and $k_Z = 10\ka/\chor$ where the temperature is larger than $ 
\kk/2\pi = 1.5 \TB$. 
The fact that this resonance occurs for wavelengths $\sim d_\xi$, 
reinforces the fact 
that the effective 
width of the horizon is indeed $d_\xi$.

\section{Conclusions}
\label{sec:conclusion}

By considering series of supersonic flows 
consisting on local perturbations $\wt$ defined on the top
of a background flow $\wB$,
we were able to identify the parameters that fix
the spectral properties of Bogoliubov phonons emitted by a (black hole)
sonic horizon.


We show that the flux remains Planckian to a high accuracy, see Eq.~\eqref{eq:run2},
even when the temperature strongly differs from $\Th = \kk/2\pi$, the standard value fixed by the surface gravity, 
see Fig.~\ref{fig:tom_fom}.  
This result implies two things.  In spite of the fact that each mode of frequency $\om$ 
has its own turning point of Eq. (\ref{eq:tp}), the Planckianity first implies that
 the temperature function $T_\om$ of Eq. (\ref{eq:Tom}) is {\it not}
determined by some function of $w(x)$ evaluated at a 
$\om$-dependent location, as $T_\om$ is common to all low frequencies with respect to $\omm$ of Eq. (\ref{eq:omm}). 
Second, it implies that the parameters governing the small deviations from Planckianity 
differ from those governing the difference between the 
low frequency temperature $T_0$ and $\Th$. 
In fact, whereas the difference $T_0 - \Th$ is governed
by the {\it near horizon} properties, deviations from
Planckianity are governed by inverse powers of $\omm/T_0$,
where $\omm$ 
is fixed by 
%
$D_{\rm as} = D + \Dt$, characterizing 
the {\it global} properties of the supersonic transition region, such as the asymptotic
velocity excess. 

Moreover, in contradistinction 
to the relativistic case
where the temperature is locally fixed by $\kk$ of Eq.~(\ref{eq:temprel}), 
we show that
the temperature $T_0$ 
is determined, to a high accuracy when $\omm/\Th \gtrsim 10$, 
by a {\it nonlocal} quantity: 
the 
spatial
average across the horizon of 
$\partial_x w$ over a width $d_\xi$, as if the horizon were
broadened.~\footnote{This is 
in agreement with what was found in~\cite{AC}
when considering the black hole laser effect~\cite{CJlaser}. In that case,
the effect disappears when the distance $\delta_{\rm H}$
between the black and the white horizon becomes too small, because
the dispersive field is unable to `resolve' the two horizons.
When using linear profiles for both horizons, one finds
that the critical value of $\delta_{\rm H} \simeq 
d_\xi$.}
%
Hence,
whenever $w(x)$ hardly varies over $d_\xi$ 
in the near horizon region,
$\bar \kappa$ of Eq.~\eqref{eq:kav}
equals the surface gravity
$\kk$, and the standard temperature is found.
Instead, when $w$ varies on scales shorter than $d_\xi$,
$T_0$
 is no longer the standard one,
but it is 
well approximated by $\bar \kappa/2\pi$ for all frequencies $\om \ll \omm$. 
This result is 
reinforced by the fact 
that the effective horizon width $d_\xi$ scales according to
a well-defined law given in
Eq.~\eqref{eq:h}.


Furthermore, 
the above 
results are basically valid for all flows. They 
cease to be valid 
under the same restricted 
conditions, namely 
when the flow contains a sharp perturbation well localized far enough from the horizon.  
Then the perturbation induces a scattering that interferes with the Hawking effect, thereby 
engendering oscillations. 
These 
are similar 
to those 
found in inflationary 
spectra, and are also attributable to interferences between two 
scatterings. 
It is clear that the local average of Eq.~\eqref{eq:kav} 
cannot describe these oscillations which are 
{nonlocal} on a scale much larger than the width $d_\xi$. 

Hence we can conclude that when $\omm/\Th \gtrsim 10 $ 
and when the 
scattering essentially arises near the horizon,
the spectrum is near Planckian, 
and
with a temperature approximately
given by 
$\bar \kappa$
of Eq.~\eqref{eq:kav}.

Finally, to mimic what is found in white hole flows~\cite{Carusottowhite,Silke},
we 
studied profiles containing an 
undulation 
on the top of a smooth profile. 
By varying its wavelength, 
we have found three different behaviors that match the above analysis.

\begin{acknowledgments} 
We are grateful to Antonin Coutant and Stefano Liberati for many discussions that took
place during this project. We thank Roberto Balbinot,
Ted Jacobson, and Stefano Liberati for helpful comments on our 
manuscript. 
We are also grateful to Jean Macher for useful explanations
concerning the code he wrote for~\cite{MacherBEC}. 
\end{acknowledgments}


\appendix

\section{
Asymptotic modes and turning points} 
\label{sec:disprel}
\begin{figure*}
 \includegraphics{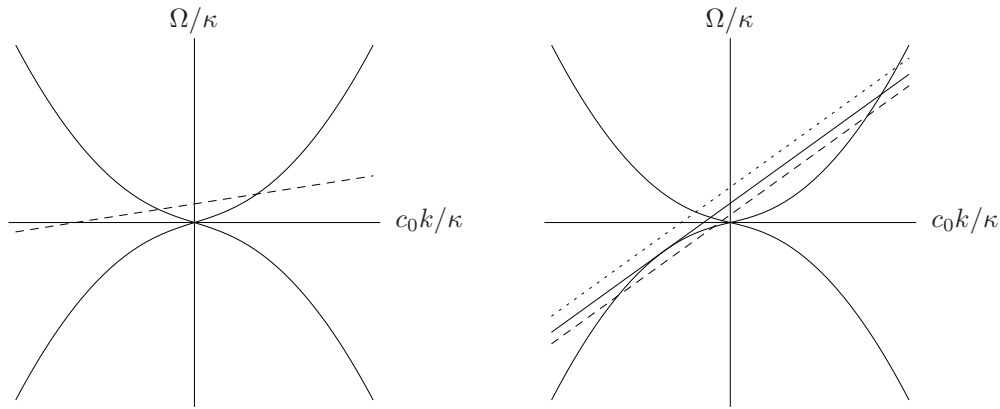}
 \caption{Graphical solution of the dispersion relation~\eqref{eq:dispersion_hom} for subsonic (left panel) and supersonic flow (right panel). Solid curves: $\pm  \Omega(k)/\kappa$.
Left panel, dashed line: $(\om-vk)/\kappa$. Right panel, dashed, solid, 
dotted lines: $\om-vk$, respectively, for $\om<\omm$, $\om=\omm$, $\om>\omm$. For $\om<\omm$,
two extra (real) roots exist in the left lower quadrant.
They cease to exist for $\om > \omm$ and this explains why there is no Hawking radiation
above $\omm$.}
 \label{fig:dispersion}
\end{figure*}


We briefly show why and how the {\it number} of modes of frequency $\om$
depends on the asymptotic properties of $w = c + v$. This is important as 
it explains why, when there is dispersion,
there is no flux above a critical frequency $\omm$ whose value is fixed
by the asymptotic value $w$.

In homogeneous subsonic flows, $|v|< c$, 
for real $\omega$, two roots $k_\om$ of Eq.~\eqref{eq:dispersion_hom} are real, 
and describe the right and left moving modes. 
The other two roots are complex, conjugated to each other, and 
describe 
 modes that are asymptotically growing or decaying, say to left.
Only the first two correspond to 
perturbations that must be quantized, as the last two are not asymptotically bound.
In infinite condensates, when canonically quantizing a field, 
only asymptotically bound modes (ABM)
should 
be included in the spectrum~\cite{MacherBEC,AC}.

In homogeneous supersonic flows, the situation is quite different.
For real $\omega$, there exists a critical frequency $\omm$ 
below which the four roots are real, as can be seen in Fig.~\ref{fig:dispersion}
(right panel), and above which two are real and two complex, as in subsonic flows.
The spectrum of ABM thus contains four modes for frequencies $0 < \om < \omm$ 
and only two for $\om > \omm$.
At $\om = \omm$ the two extra roots merge and the curve $\Omega(k)$
is tangent to the line $\om - vk$. 
Hence 
the group velocity 
in the lab
\begin{equation}\label{eq:gr_vel}
v_{\rm gr} = 
\partial_k \omega =
\partial_k (\Omega + vk), 
\end{equation}
vanishes at that frequency.
Solving this equation one finds that $\omm$ is
proportional to $\Lambda$, but also depends also on the supersonic velocity excess $|v| - c$
(see Fig.~4 in~\cite{MacherRP1} for details).

%
%
%

In 
{\it inhomogeneous} flows that cross once the speed of sound  
and with two asymptotic flat regions, i.e., as those of Eq.~\eqref{eq:velocityB},
one gets a new 
situation: 
when $\om>\omm$,
there are still 
only two modes, 
whereas for $\om<\omm$, 
 there are now three independent 
ABM~\cite{MacherRP1}.  
There are only three independent modes because
the semiclassical trajectories associated with the two extra roots  
necessarily possess a turning point for $x < 0$, in the 
supersonic region. Hence these two roots describe the initial and final value of the momentum
of the same mode. 

The location of the
turning point $x^{\rm tp}$ depends on $\om$. Its value is easily obtains by
solving $v_{\rm gr} = 0$ [see Eq.~\eqref{eq:gr_vel}].
Using Eqs.~\eqref{eq:temprel} and~\eqref{eq:dispersion_hom}, 
for $\om \ll \Lambda$, one 
finds~\cite{CJ96}
%
\begin{equation}
\label{eq:tp}
-\frac{\kk  x^{\rm tp}(\om)}{c_0} \approx \left(\frac{\om}{\Lambda}\right)^{2/3}.
\end{equation}
For $\om \to 0$ fixed $\Lambda$, or $\Lambda \to \infty$ fixed $\om$,
the turning point coincides with the Killing horizon at $x=0$, as expected
from the behavior of the semiclassical trajectories in the absence of dispersion.
[When considering subluminal dispersion by flipping the sign of the $k^4$ term in Eq.~\eqref{eq:dispersion_hom},
one verifies that $x^{\rm tp}(\om)$ flips sign but keeps the same norm.] 

When $\om \to \omm$, the turning point $x^{\rm tp}(\om) \to -\infty$.
Hence, 
the threshold frequency
$\omm$ is now fixed by the {\it asymptotic} value of $w$ for $x \to - \infty$.
In our velocity flows of Eq.~\eqref{eq:velocity_dec}, one has
\begin{equation}\label{eq:omm}
\omm = \Lambda \, f(D_{\rm as}),
\end{equation}
%
where $D_{\rm as} = -w(x = - \infty)$.
Using Eqs.~\eqref{eq:velocityB} and~\eqref{eq:velocity2}, 
 one has $D_{\rm as} =D + \Dt$.
%

\section{Mode analysis and scattering matrix} 
\label{sec:bogo}

We 
present the main concepts that characterize the 
scattering of phonons in atomic Bose condensates
in stationary flows containing one sonic horizon. We follow the treatment of Ref.~\cite{MacherBEC}
where
more details can be found.

Phonon elementary excitations are described by a quantum field operator, $\hp$. 
In this work, 
we shall use 
the {\it relative} perturbations defined by
%
\begin{equation}
 \hP=\Pn(1+\hp),
\end{equation}
where $\hP$ is the atom's field operator, and  $\Pn$ the condensate. 
The dynamics of $\hp$ is determined by the Bogoliubov--de~Gennes equation~\cite{DalfovoRMP}.
When using $\hp$ and when $\hbar = 1$, 
one obtains 
\begin{equation}\label{eq:bdg}
  \ii 
\pdt\hp = \left[T_\rho-\ii v 
\pdx + m c^2 \right] \hp + m c^2 \hpd,
\end{equation}
where
$c$ is the $x$-dependent speed of sound
\begin{equation}
 c^2(x)\equiv\frac{g(x)\rn(x)}{m},
\end{equation}
where $\rn(x)= |\Pn(x) |^2$ is the mean density of condensed atoms,
and $g(x)$ is the effective coupling constant among atoms. These
functions play no role in the sequel: only $c(x)$ and $v(x)$ matter. The quantity
%
\begin{equation}\label{eq:Trho}
 T_\rho
=-\frac{ 
1 }{2m} \, v\pdx \frac{1}{v}\pdx,
\end{equation}
is the kinetic operator that acts on $\hp$.

In stationary situations, 
$\hp$ can be expanded in Killing frequency eigenmodes
\begin{multline}\label{eq:phiexpansion}
\hp =  \int\!\dom\left[\phm{\om t}\hp_\om(x)+\php{\om t} \hat \varphi_\om(x)^\dagger\right] \\ 
=\int\!\dom\sum_{\alpha}\left[\phm{\om t}\pom\aom+\php{\om t}\vpoms\aomd\right],
\end{multline}
where the discrete sum over $\alpha$ 
takes into account the number of ABM at each $\om$.
Inserting Eq.~\eqref{eq:phiexpansion} in Eq.~\eqref{eq:bdg} yields the system 
for the couple of modes $\pom, \vpom$ 
\begin{equation}\label{eq:system}
\begin{aligned}
 \left[ 
(\om +\ii v\pdx) - T_\rho - mc^2\right]\pom=  mc^2\vpom,\\
 \left[-
(\om +\ii v\pdx) - T_\rho - mc^2\right]\vpom = mc^2\pom.
 \end{aligned}
\end{equation}

%

As discussed above, 
in backgrounds containing one sonic horizon, 
they are two or three ABM depending on
whether $\om$ is larger or smaller than $\omm$.
%
%
Therefore the $\om$ component of the field operator 
reads 
\begin{equation}\label{eq:planewaves}
 \hp_\om =\phi_\om^u \ha_\om^u+\phi_\om^v\ha_\om^v
+ \theta(\ommax - \om) \varphi_{-\om}^*\ha_{-\om}^{
\dagger}.
\end{equation}
%
The modes $\phi_\om^u$ and $\phi_\om^v$ have positive norm, they 
are associated with the usual
real roots of Eq.~\eqref{eq:dispersion_hom} (see Fig.~\ref{fig:dispersion}), and they describe, 
respectively, 
propagating right-$u$ and left-$v$ moving waves with respect to the condensed atoms. 
On the contrary, $(\varphi_{-\om})^*$ 
has a negative norm. It is associated with the extra roots
that exist in supersonic flows for $\om < \omm$, and 
describes 
phonons that are trapped inside the horizon ($x<0$)
in the supersonic region. 
These modes describe
the negative frequency partners of the outgoing Hawking 
phonons
represented
by $\phi_\om^u$ and $\ha_\om^u$.

In our 
infinite condensates, 
the above modes become superpositions of 
plane waves both in the left and right asymptotic regions of Eq.~\eqref{eq:planewaves}.
We can thus construct without ambiguity the in- and out-mode bases.
The in modes are such that 
each of them contains 
only one asymptotic branch 
with 
group velocity directed towards the horizon. 
The definition of out modes is analogous and based on the criterion that 
each out mode contains 
only one asymptotic branch 
with group velocity directed away from the horizon. 


As shown in Appendix~\ref{sec:disprel}, 
for $\om>\omm$, 
there exist only two positive norm modes, and therefore in- and out-mode bases are 
linearly related by a trivial (elastic) $2 \times 2$ transformation. 
Instead, when  $\om < \omm$, the in and out bases contain three modes
which mix with each other in a nontrivial way by a  
$3 \times 3 $ 
Bogoliubov transformation~\cite{MacherBEC}
\begin{equation}\label{eq:bog_transf}
\begin{aligned}
 \phi_\om^{u,{\rm in}} &= \alpha_\om \phi_\om^{u,\rm out} + \beta_{-\om} \left(\varphi_{-\om}^{\rm out}\right)^*
 				+ \tilde A_\om \phi_\om^{v,\rm out},\\
 (\varphi_{-\om}^{
{\rm in}})^* &= \beta_\om 
\phi_\om^{u,\rm out} 
+ \alpha_{-\om} \left(\varphi_{-\om}^{\rm out}\right)^* 
 				+ \tilde B_\om 
\phi_\om^{v,\rm out} 
,\\
 \phi_\om^{v,{\rm in}} &= A_\om \phi_\om^{u,\rm out} + B_{\om} \left(\varphi_{-\om}^{\rm out}\right)^*
 				+ \alpha_\om^v \phi_\om^{v,\rm out}.
\end{aligned}
\end{equation}
The standard 
normalization of the modes yields relations such as (from the first equation)
\begin{equation}
 |\alpha_\om|^2 - |\beta_{-\om}|^2 + |\tilde A_\om|^2 = 1.
\end{equation}
%
When the initial state is vacuum, the final occupation numbers are 
given by the Bogoliubov coefficients
\begin{equation}
\begin{gathered}
 n_\om 
= |\beta_\om|^2,\quad
  n_\om^{v} 
= |\tilde B_\om|^2,\\
  n_{-\om} 
 = 
   n_\om +  n_\om^{v}.
\end{gathered}
\end{equation}
From these relations one sees 
that the norm of $\beta_\om$ 
fixes 
the occupation number of outgoing phonons 
spontaneously produced by the scattering of the
quantum field near the 
black hole horizon. In the standard analysis without dispersion,
$n_\om$ is Planckian and describes the Hawking effect~\cite{Primer}.
In that case, for massless conformally invariant fields, 
$ n_\om^{v}$, the number of left moving phonons spontaneously produced, identically vanishes.
In the presence of dispersion, one generally finds $ n_\om^{v} \ll n_\om$. Hence the scattering
of a dispersive field on a sonic horizon basically consists of a more general and 
slightly modified version of the standard Hawking effect. This is true when 
$\omm \gtrsim 2 \kappa$. For 
larger values of $\kappa$,
the deviations become large, but the 
structure of Eq.~\eqref{eq:bog_transf} and the meaning of its coefficients remain 
unchanged~\cite{MacherBEC,Recati2009}.
\bibliography{rob}

\begin{thebibliography}{31}
\expandafter\ifx\csname natexlab\endcsname\relax\def\natexlab#1{#1}\fi
\expandafter\ifx\csname bibnamefont\endcsname\relax
  \def\bibnamefont#1{#1}\fi
\expandafter\ifx\csname bibfnamefont\endcsname\relax
  \def\bibfnamefont#1{#1}\fi
\expandafter\ifx\csname citenamefont\endcsname\relax
  \def\citenamefont#1{#1}\fi
\expandafter\ifx\csname url\endcsname\relax
  \def\url#1{\texttt{#1}}\fi
\expandafter\ifx\csname urlprefix\endcsname\relax\def\urlprefix{URL }\fi
\providecommand{\bibinfo}[2]{#2}
\providecommand{\eprint}[2][]{\url{#2}}

\bibitem[{\citenamefont{Unruh}(1981)}]{Unruh81}
\bibinfo{author}{\bibfnamefont{W.~G.} \bibnamefont{Unruh}},
  \bibinfo{journal}{Phys. Rev. Lett.} \textbf{\bibinfo{volume}{46}},
  \bibinfo{pages}{1351} (\bibinfo{year}{1981}).

\bibitem[{\citenamefont{Jacobson}(1991)}]{TJ91}
\bibinfo{author}{\bibfnamefont{T.}~\bibnamefont{Jacobson}},
  \bibinfo{journal}{Phys. Rev. D} \textbf{\bibinfo{volume}{44}},
  \bibinfo{pages}{1731} (\bibinfo{year}{1991}).

\bibitem[{\citenamefont{Jacobson}(1993)}]{93}
\bibinfo{author}{\bibfnamefont{T.}~\bibnamefont{Jacobson}},
  \bibinfo{journal}{Phys. Rev. D} \textbf{\bibinfo{volume}{48}},
  \bibinfo{pages}{728} (\bibinfo{year}{1993}).

\bibitem[{\citenamefont{Brout et~al.}(1995{\natexlab{a}})\citenamefont{Brout,
  Massar, Parentani, and Spindel}}]{Primer}
\bibinfo{author}{\bibfnamefont{R.}~\bibnamefont{Brout}},
  \bibinfo{author}{\bibfnamefont{S.}~\bibnamefont{Massar}},
  \bibinfo{author}{\bibfnamefont{R.}~\bibnamefont{Parentani}},
  \bibnamefont{and} \bibinfo{author}{\bibfnamefont{P.}~\bibnamefont{Spindel}},
  \bibinfo{journal}{Phys. Rept.} \textbf{\bibinfo{volume}{260}},
  \bibinfo{pages}{329} (\bibinfo{year}{1995}{\natexlab{a}}).

\bibitem[{\citenamefont{Unruh}(1995)}]{Unruh95}
\bibinfo{author}{\bibfnamefont{W.~G.} \bibnamefont{Unruh}},
  \bibinfo{journal}{Phys. Rev. D} \textbf{\bibinfo{volume}{51}},
  \bibinfo{pages}{2827} (\bibinfo{year}{1995}).

\bibitem[{\citenamefont{Brout et~al.}(1995{\natexlab{b}})\citenamefont{Brout,
  Massar, Parentani, and Spindel}}]{BMPS95}
\bibinfo{author}{\bibfnamefont{R.}~\bibnamefont{Brout}},
  \bibinfo{author}{\bibfnamefont{S.}~\bibnamefont{Massar}},
  \bibinfo{author}{\bibfnamefont{R.}~\bibnamefont{Parentani}},
  \bibnamefont{and} \bibinfo{author}{\bibfnamefont{P.}~\bibnamefont{Spindel}},
  \bibinfo{journal}{Phys. Rev. D} \textbf{\bibinfo{volume}{52}},
  \bibinfo{pages}{4559} (\bibinfo{year}{1995}{\natexlab{b}}).

\bibitem[{\citenamefont{Corley}(1998)}]{Corley97}
\bibinfo{author}{\bibfnamefont{S.}~\bibnamefont{Corley}},
  \bibinfo{journal}{Phys. Rev. D} \textbf{\bibinfo{volume}{57}},
  \bibinfo{pages}{6280} (\bibinfo{year}{1998}).

\bibitem[{\citenamefont{Himemoto and Tanaka}(2000)}]{Tanaka99}
\bibinfo{author}{\bibfnamefont{Y.}~\bibnamefont{Himemoto}} \bibnamefont{and}
  \bibinfo{author}{\bibfnamefont{T.}~\bibnamefont{Tanaka}},
  \bibinfo{journal}{Phys. Rev. D} \textbf{\bibinfo{volume}{61}},
  \bibinfo{pages}{064004} (\bibinfo{year}{2000}).

\bibitem[{\citenamefont{Unruh and Schutzhold}(2005)}]{UnruhSchu05}
\bibinfo{author}{\bibfnamefont{W.~G.} \bibnamefont{Unruh}} \bibnamefont{and}
  \bibinfo{author}{\bibfnamefont{R.}~\bibnamefont{Schutzhold}},
  \bibinfo{journal}{Phys. Rev. D} \textbf{\bibinfo{volume}{71}},
  \bibinfo{pages}{024028} (\bibinfo{year}{2005}).

\bibitem[{\citenamefont{Corley and Jacobson}(1996)}]{CJ96}
\bibinfo{author}{\bibfnamefont{S.}~\bibnamefont{Corley}} \bibnamefont{and}
  \bibinfo{author}{\bibfnamefont{T.}~\bibnamefont{Jacobson}},
  \bibinfo{journal}{Phys. Rev. D} \textbf{\bibinfo{volume}{54}},
  \bibinfo{pages}{1568} (\bibinfo{year}{1996}).

\bibitem[{\citenamefont{Unruh}(2007)}]{UnruhTrieste}
\bibinfo{author}{\bibfnamefont{W.~G.} \bibnamefont{Unruh}},
  \bibinfo{journal}{PoS} \textbf{\bibinfo{volume}{QG-PH}}, \bibinfo{pages}{039}
  (\bibinfo{year}{2007}).

\bibitem[{\citenamefont{Macher and Parentani}(2009)}]{MacherRP1}
\bibinfo{author}{\bibfnamefont{J.}~\bibnamefont{Macher}} \bibnamefont{and}
  \bibinfo{author}{\bibfnamefont{R.}~\bibnamefont{Parentani}},
  \bibinfo{journal}{Phys. Rev. D} \textbf{\bibinfo{volume}{79}},
  \bibinfo{pages}{124008} (\bibinfo{year}{2009}).

\bibitem[{\citenamefont{{Macher} and {Parentani}}(2009)}]{MacherBEC}
\bibinfo{author}{\bibfnamefont{J.}~\bibnamefont{{Macher}}} \bibnamefont{and}
  \bibinfo{author}{\bibfnamefont{R.}~\bibnamefont{{Parentani}}},
  \bibinfo{journal}{Phys. Rev. A} \textbf{\bibinfo{volume}{80}},
  \bibinfo{pages}{043601} (\bibinfo{year}{2009}).

\bibitem[{\citenamefont{Coutant and Parentani}()}]{ACRP2}
\bibinfo{author}{\bibfnamefont{A.}~\bibnamefont{Coutant}} \bibnamefont{and}
  \bibinfo{author}{\bibfnamefont{R.}~\bibnamefont{Parentani}},
  \bibinfo{journal}{unpublished}.

\bibitem[{\citenamefont{{Mayoral} et~al.}(2011)\citenamefont{{Mayoral},
  {Recati}, {Fabbri}, {Parentani}, {Balbinot}, and
  {Carusotto}}}]{Carusottowhite}
\bibinfo{author}{\bibfnamefont{C.}~\bibnamefont{{Mayoral}}},
  \bibinfo{author}{\bibfnamefont{A.}~\bibnamefont{{Recati}}},
  \bibinfo{author}{\bibfnamefont{A.}~\bibnamefont{{Fabbri}}},
  \bibinfo{author}{\bibfnamefont{R.}~\bibnamefont{{Parentani}}},
  \bibinfo{author}{\bibfnamefont{R.}~\bibnamefont{{Balbinot}}},
  \bibnamefont{and}
  \bibinfo{author}{\bibfnamefont{I.}~\bibnamefont{{Carusotto}}},
  \bibinfo{journal}{New J. Phys.} \textbf{\bibinfo{volume}{13}},
  \bibinfo{pages}{025007} (\bibinfo{year}{2011}).

\bibitem[{\citenamefont{{Weinfurtner} et~al.}(2011)\citenamefont{{Weinfurtner},
  {Tedford}, {Penrice}, {Unruh}, and {Lawrence}}}]{Silke}
\bibinfo{author}{\bibfnamefont{S.}~\bibnamefont{{Weinfurtner}}},
  \bibinfo{author}{\bibfnamefont{E.~W.} \bibnamefont{{Tedford}}},
  \bibinfo{author}{\bibfnamefont{M.~C.~J.} \bibnamefont{{Penrice}}},
  \bibinfo{author}{\bibfnamefont{W.~G.} \bibnamefont{{Unruh}}},
  \bibnamefont{and} \bibinfo{author}{\bibfnamefont{G.~A.}
  \bibnamefont{{Lawrence}}}, \bibinfo{journal}{Phys. Rev. Lett.}
  \textbf{\bibinfo{volume}{106}}, \bibinfo{pages}{021302}
  (\bibinfo{year}{2011}).

\bibitem[{\citenamefont{Barcelo et~al.}(2005)\citenamefont{Barcelo, Liberati,
  and Visser}}]{lr}
\bibinfo{author}{\bibfnamefont{C.}~\bibnamefont{Barcelo}},
  \bibinfo{author}{\bibfnamefont{S.}~\bibnamefont{Liberati}}, \bibnamefont{and}
  \bibinfo{author}{\bibfnamefont{M.}~\bibnamefont{Visser}},
  \bibinfo{journal}{Living Rev. Rel.} \textbf{\bibinfo{volume}{8}},
  \bibinfo{pages}{12} (\bibinfo{year}{2005}).

\bibitem[{\citenamefont{Bardeen et~al.}(1973)\citenamefont{Bardeen, Carter, and
  Hawking}}]{BCH}
\bibinfo{author}{\bibfnamefont{J.~M.} \bibnamefont{Bardeen}},
  \bibinfo{author}{\bibfnamefont{B.}~\bibnamefont{Carter}}, \bibnamefont{and}
  \bibinfo{author}{\bibfnamefont{S.~W.} \bibnamefont{Hawking}},
  \bibinfo{journal}{Commun. Math. Phys.} \textbf{\bibinfo{volume}{31}},
  \bibinfo{pages}{161} (\bibinfo{year}{1973}).

\bibitem[{\citenamefont{{Misner} et~al.}(1973)\citenamefont{{Misner}, {Thorne},
  and {Wheeler}}}]{MisnerThorneWheeler}
\bibinfo{author}{\bibfnamefont{C.~W.} \bibnamefont{{Misner}}},
  \bibinfo{author}{\bibfnamefont{K.~S.} \bibnamefont{{Thorne}}},
  \bibnamefont{and} \bibinfo{author}{\bibfnamefont{J.~A.}
  \bibnamefont{{Wheeler}}}, \emph{\bibinfo{title}{{Gravitation}}}
  (\bibinfo{publisher}{{W.H.~Freeman and Co.}}, \bibinfo{address}{{San
  Francisco}}, \bibinfo{year}{1973}).

\bibitem[{\citenamefont{Balbinot et~al.}(2005)\citenamefont{Balbinot, Fabbri,
  Fagnocchi, and Parentani}}]{Rivista}
\bibinfo{author}{\bibfnamefont{R.}~\bibnamefont{Balbinot}},
  \bibinfo{author}{\bibfnamefont{A.}~\bibnamefont{Fabbri}},
  \bibinfo{author}{\bibfnamefont{S.}~\bibnamefont{Fagnocchi}},
  \bibnamefont{and}
  \bibinfo{author}{\bibfnamefont{R.}~\bibnamefont{Parentani}},
  \bibinfo{journal}{Riv. Nuovo Cim.} \textbf{\bibinfo{volume}{28}},
  \bibinfo{pages}{1} (\bibinfo{year}{2005}).

\bibitem[{\citenamefont{Parentani}(2010)}]{From2010}
\bibinfo{author}{\bibfnamefont{R.}~\bibnamefont{Parentani}},
  \bibinfo{journal}{Phys. Rev. D} \textbf{\bibinfo{volume}{82}},
  \bibinfo{pages}{025008} (\bibinfo{year}{2010}).

\bibitem[{\citenamefont{Dalfovo et~al.}(1999)\citenamefont{Dalfovo, Giorgini,
  Pitaevskii, and Stringari}}]{DalfovoRMP}
\bibinfo{author}{\bibfnamefont{F.}~\bibnamefont{Dalfovo}},
  \bibinfo{author}{\bibfnamefont{S.}~\bibnamefont{Giorgini}},
  \bibinfo{author}{\bibfnamefont{L.~P.} \bibnamefont{Pitaevskii}},
  \bibnamefont{and}
  \bibinfo{author}{\bibfnamefont{S.}~\bibnamefont{Stringari}},
  \bibinfo{journal}{Rev. Mod. Phys.} \textbf{\bibinfo{volume}{71}},
  \bibinfo{pages}{463} (\bibinfo{year}{1999}).

\bibitem[{\citenamefont{Schutzhold and Unruh}(2008)}]{08}
\bibinfo{author}{\bibfnamefont{R.}~\bibnamefont{Schutzhold}} \bibnamefont{and}
  \bibinfo{author}{\bibfnamefont{W.~G.} \bibnamefont{Unruh}},
  \bibinfo{journal}{Phys. Rev. D} \textbf{\bibinfo{volume}{78}},
  \bibinfo{pages}{041504} (\bibinfo{year}{2008}).

\bibitem[{\citenamefont{Parentani}(2001)}]{quantum_metric_fluctRP}
\bibinfo{author}{\bibfnamefont{R.}~\bibnamefont{Parentani}},
  \bibinfo{journal}{Phys. Rev. D} \textbf{\bibinfo{volume}{63}},
  \bibinfo{pages}{041503} (\bibinfo{year}{2001}).

\bibitem[{\citenamefont{Parentani}(2002)}]{Beyond_RP}
\bibinfo{author}{\bibfnamefont{R.}~\bibnamefont{Parentani}},
  \bibinfo{journal}{Int. J. Theor. Phys.} \textbf{\bibinfo{volume}{41}},
  \bibinfo{pages}{2175} (\bibinfo{year}{2002}).

\bibitem[{\citenamefont{Jacobson and Parentani}(2007)}]{ent_ent_TJRP}
\bibinfo{author}{\bibfnamefont{T.}~\bibnamefont{Jacobson}} \bibnamefont{and}
  \bibinfo{author}{\bibfnamefont{R.}~\bibnamefont{Parentani}},
  \bibinfo{journal}{Phys. Rev. D} \textbf{\bibinfo{volume}{76}},
  \bibinfo{pages}{024006} (\bibinfo{year}{2007}).

\bibitem[{\citenamefont{Campo et~al.}(2007)\citenamefont{Campo, Niemeyer, and
  Parentani}}]{CNP}
\bibinfo{author}{\bibfnamefont{D.}~\bibnamefont{Campo}},
  \bibinfo{author}{\bibfnamefont{J.~C.} \bibnamefont{Niemeyer}},
  \bibnamefont{and}
  \bibinfo{author}{\bibfnamefont{R.}~\bibnamefont{Parentani}},
  \bibinfo{journal}{Phys. Rev. D} \textbf{\bibinfo{volume}{76}},
  \bibinfo{pages}{023513} (\bibinfo{year}{2007}).

\bibitem[{\citenamefont{Corley and Jacobson}(1999)}]{CJlaser}
\bibinfo{author}{\bibfnamefont{S.}~\bibnamefont{Corley}} \bibnamefont{and}
  \bibinfo{author}{\bibfnamefont{T.}~\bibnamefont{Jacobson}},
  \bibinfo{journal}{Phys. Rev. D} \textbf{\bibinfo{volume}{59}},
  \bibinfo{pages}{124011} (\bibinfo{year}{1999}).

\bibitem[{\citenamefont{Coutant and Parentani}(2010)}]{AC}
\bibinfo{author}{\bibfnamefont{A.}~\bibnamefont{Coutant}} \bibnamefont{and}
  \bibinfo{author}{\bibfnamefont{R.}~\bibnamefont{Parentani}},
  \bibinfo{journal}{Phys. Rev. D} \textbf{\bibinfo{volume}{81}},
  \bibinfo{pages}{084042} (\bibinfo{year}{2010}).

\bibitem[{\citenamefont{Finazzi and Parentani}(2010)}]{bhlasers}
\bibinfo{author}{\bibfnamefont{S.}~\bibnamefont{Finazzi}} \bibnamefont{and}
  \bibinfo{author}{\bibfnamefont{R.}~\bibnamefont{Parentani}},
  \bibinfo{journal}{New J. Phys.} \textbf{\bibinfo{volume}{12}},
  \bibinfo{pages}{095015} (\bibinfo{year}{2010}).

\bibitem[{\citenamefont{Recati et~al.}(2009)\citenamefont{Recati, Pavloff, and
  Carusotto}}]{Recati2009}
\bibinfo{author}{\bibfnamefont{A.}~\bibnamefont{Recati}},
  \bibinfo{author}{\bibfnamefont{N.}~\bibnamefont{Pavloff}}, \bibnamefont{and}
  \bibinfo{author}{\bibfnamefont{I.}~\bibnamefont{Carusotto}},
  \bibinfo{journal}{Phys. Rev. A} \textbf{\bibinfo{volume}{80}},
  \bibinfo{pages}{043603} (\bibinfo{year}{2009}), \eprint{0907.4305}.

\end{thebibliography}

\end{document}